\documentclass[journal]{IEEEtran}
\usepackage[colorlinks]{hyperref}

\ifCLASSINFOpdf
\else
\fi

\usepackage{graphics}
\usepackage{tikz}

\usetikzlibrary{matrix,positioning}
\usepackage{graphicx} 

\usepackage{amsmath}
\usepackage{mathtools}
\usepackage{cleveref}

\graphicspath{{../cc_review/Figures/}}
\usetikzlibrary{arrows,shapes,positioning,shadows,trees}
\usepackage{pgfplots}
\pgfplotsset{compat = newest}

\tikzset{
	basic/.style  = {draw, text width=2cm, drop shadow, font=\sffamily, rectangle},
	root/.style   = {basic, rounded corners=2pt, thin, align=center,
		fill=green!30},
	level 2/.style = {basic, rounded corners=6pt, thin,align=center, fill=green!60,
		text width=8em},
	level 3/.style = {basic, thin, align=left, fill=pink!60, text width=6.5em}
}

\usetikzlibrary{positioning,fit}
\tikzstyle{road-draw}=[draw=black,thick]
\tikzstyle{road-fill}=[black!10]

\tikzstyle{message}=[fill=white,draw=black,thick]

\tikzstyle{wave-draw}=[draw=blue!50!black,dotted,very thick]
\tikzstyle{wave-fill}=[fill=blue!50,opacity=.2]
\tikzstyle{accident-star}=[draw=red,fill=red!60,star,star points=10]
\usetikzlibrary{%
	chains,
	calc,
	shapes,
	decorations,
	scopes,
	arrows,
	svg.path,
}

\tikzstyle{lane-right}=[start chain=going right,node distance=1ex]
\tikzstyle{lane-left}=[start chain=going left,node distance=1ex]

\newcommand\basicroad{%
\foreach \pos/\coord in {top/3ex, center/0, bottom/-3ex}{
\coordinate (\pos-left) at (0,\coord);
\coordinate (\pos-right) at (\textwidth-\pgflinewidth,\coord);
}

\coordinate (lane1-left) at (1,-1.5ex);
\path (\textwidth-\pgflinewidth,-1.5ex) +(-1,0) coordinate (lane1-right);
\coordinate (lane2-left) at (1,1.5ex);
\path (\textwidth-\pgflinewidth,1.5ex) +(-1,0) coordinate (lane2-right);

\coordinate (lane1) at (lane1-left);
\coordinate (lane2) at (lane2-right);

\fill[road-fill] (top-left) rectangle (bottom-right);
\draw[road-draw] (top-left) -- (top-right);
\draw[road-draw,loosely dashed] (center-left) -- (center-right);
}

\newcommand\onramproad{%
\basicroad

\fill[road-fill] (bottom-left) -- ++(3,0) to[out=0,in=180] ++(1,-3ex) -- ++(3,0) coordinate (bend) to[out=0,in=135] ++(1,-3ex) -- ++(45:3ex) to[out=135,in=0] (bend |- bottom-left) -- cycle;

\draw[road-draw] (bottom-left) -- ++(3,0) coordinate (ramp-start) to[out=0,in=180] ++(1,-3ex) -- ++(3,0) coordinate (bend) to[out=0,in=135] ++(1,-3ex) ++(45:3ex) to[out=135,in=0] (bend |- bottom-left) coordinate (ramp-end) -- (bottom-right);

\draw[road-draw,dashed] (ramp-start) -- (ramp-end);
}

\tikzset{message/.style={->,shorten >=2pt,shorten <=2pt,line width=1pt}}

\newcommand\commwave[2]{%
\path[wave-fill] #1 circle (#2);
\path[wave-draw] #1 circle (#2);
}

\newcommand\accident[2]{%
\path (lane1-#1) +(#2,0) coordinate (p1);
\path (lane2-#1) +(#2,0) coordinate (p2);

\path (p1) edge node[right,accident-star,minimum size=4ex] {} (p2);
\node[Car,rotate=30] at (p1) {};
\node[Car,rotate=-50] at (p2) {};
}

\tikzset{Car/.style={path picture={
\pgftransformscale{1em/10mm}
\fill[#1] svg "M28.21-13.605c9.562,3.446,9.648,23.671,0.155,27.315 c-4.363,0.507-11.834,0.663-14.102-0.052l-15.559,0.098c-0.361,0.54-2.344,0.587-2.49-0.046l-9.12-0.052 c-2.42,0.596-9.124,0.491-13.695,0.354c-0.868-0.069-2.466-0.383-3.055-0.86c-0.586-0.088-2.902-0.613-3.526-2.013 c-2.881-3.816-3.105-17.391-0.049-22.328c0.849-1.417,2.686-1.826,3.369-1.86c1.172-0.46,2.07-0.8,2.747-0.914 c4.707-0.249,12.097-0.404,14.099,0.304h9.073c0.314-0.542,2.336-0.684,2.805-0.049l15.296,0.103 C16.197-14.515,25.906-14.046,28.21-13.605z";%
\fill[CarWindow] svg "M-24.197,9.442c-0.964,0.555-2.364,1.198-2.94,1.826c7.256,0.867,22.892,1.016,38.82,1.092 c-1.101-0.841-5.213-1.443-8.749-2.128C-0.618,9.9-14.507,9.656-24.197,9.442z";%
\fill[CarWindow] svg "M33.182,7.616l-1.928,0.117c-0.877,1.522-1.91,4.131-4.51,5.058 c0.353,0.272,0.898,0.205,1.379,0.233C31.92,11.97,32.73,8.522,33.182,7.616z";%
\fill[CarWindow] svg "M-31.825-9.373c-2.358,4.111-2.324,14.41,0.069,18.753l4.252-0.791 c-2.718,0.159-2.64-17.187,0.057-17.171L-31.825-9.373z";%
\fill[CarWindow] svg "M14.187-11.443c-2.099,0.058-6.936,1.246-10.511,1.89C5.151-6.242,4.517,7.381,3.746,9.752 c3.689,0.599,8.887,1.858,10.498,1.517C18.698,9.763,18.805-9.979,14.187-11.443z";%
\fill[CarWindow] svg "M11.803-12.355c-15.93,0.087-31.488,0.225-38.756,1.103c0.581,0.619,1.916,1.206,2.886,1.758 c9.685-0.21,23.573-0.4,27.118-0.732C6.597-10.911,10.71-11.511,11.803-12.355z";%
\fill[CarWindow] svg "M27.243-13.147c-0.235,0.029-0.453,0.103-0.626,0.249c2.599,0.919,3.635,3.526,4.503,5.049 l1.934,0.127c-0.438-0.917-1.317-4.36-5.118-5.424C27.691-13.128,27.474-13.162,27.243-13.147z";%
},minimum width=2.5em,minimum height=1em},
Car/.default={black},
CarWindow/.style={fill=white},
}

\tikzset{Rsu/.style={path picture={
\pgftransformscale{3em/14mm}
\pgftransformrotate{180}
\fill[#1] svg "M5.188,18.438c-0.242,0-0.454-0.175-0.493-0.422L0.697-7.22C0.654-7.492,0.84-7.749,1.113-7.792
	c0.271-0.043,0.528,0.143,0.572,0.416l3.998,25.235c0.043,0.272-0.143,0.529-0.416,0.572C5.241,18.436,5.214,18.438,5.188,18.438z";
\fill[#1] svg "M-5.624,18.74c-0.026,0-0.053-0.002-0.08-0.006c-0.272-0.044-0.458-0.301-0.414-0.573L-1.682-9.34
	c0.043-0.272,0.299-0.461,0.573-0.414c0.272,0.044,0.458,0.301,0.414,0.573L-5.131,18.32C-5.17,18.565-5.383,18.74-5.624,18.74z";
\fill[#1] svg "M-5.019,17.469c-0.163,0-0.322-0.079-0.418-0.226c-0.151-0.231-0.087-0.541,0.144-0.692l9.001-5.911
	c0.23-0.151,0.541-0.087,0.692,0.144c0.151,0.231,0.087,0.541-0.144,0.692l-9.001,5.911C-4.83,17.442-4.925,17.469-5.019,17.469z";
\fill[#1] svg "M4.583,17.263c-0.094,0-0.189-0.026-0.273-0.082l-8.649-5.663c-0.231-0.151-0.295-0.461-0.145-0.692
	c0.152-0.23,0.461-0.295,0.692-0.145l8.649,5.663c0.231,0.151,0.295,0.461,0.145,0.692C4.906,17.183,4.746,17.263,4.583,17.263z";
\fill[#1] svg "M-3.774,10.612c-0.157,0-0.312-0.074-0.409-0.211c-0.159-0.226-0.105-0.538,0.12-0.697l6.807-4.806
	C2.97,4.74,3.282,4.793,3.441,5.018c0.159,0.226,0.105,0.538-0.12,0.697l-6.807,4.806C-3.574,10.583-3.674,10.612-3.774,10.612z";
\fill[#1] svg "M3.63,10.612c-0.1,0-0.2-0.03-0.288-0.092l-6.875-4.855c-0.226-0.159-0.279-0.471-0.12-0.697
	c0.159-0.226,0.471-0.279,0.697-0.12l6.875,4.855c0.226,0.159,0.279,0.471,0.12,0.697C3.941,10.539,3.787,10.612,3.63,10.612z";
\fill[#1] svg "M-2.935,4.747c-0.143,0-0.285-0.061-0.384-0.179c-0.177-0.212-0.149-0.527,0.062-0.704l5.195-4.346
	C2.152-0.66,2.466-0.63,2.643-0.419C2.82-0.208,2.792,0.108,2.58,0.285L-2.615,4.63C-2.708,4.709-2.822,4.747-2.935,4.747z";
\fill[#1] svg "M2.792,4.747c-0.113,0-0.227-0.038-0.32-0.116l-5.198-4.347C-2.938,0.107-2.966-0.208-2.79-0.42
	c0.177-0.21,0.492-0.24,0.705-0.063l5.198,4.347c0.212,0.177,0.24,0.493,0.063,0.705C3.077,4.686,2.935,4.747,2.792,4.747z";
\fill[#1] svg "M-0.005-12.703c-1.03,0.001-1.999,0.404-2.726,1.135C-3.459-10.84-3.858-9.87-3.856-8.827
	c0.005,2.123,1.735,3.851,3.857,3.851h0.006c1.03-0.002,1.998-0.405,2.726-1.135c0.727-0.73,1.126-1.699,1.125-2.741
	C3.853-10.976,2.123-12.703-0.005-12.703z";
\fill[#1] svg "M-4.533-3.745c-0.083,0-0.168-0.021-0.247-0.065c-1.347-0.765-2.315-2.009-2.726-3.502
	c-0.411-1.494-0.215-3.057,0.552-4.404c0.492-0.865,1.201-1.591,2.05-2.1c0.237-0.142,0.543-0.065,0.686,0.171
	c0.142,0.237,0.065,0.544-0.171,0.686c-0.702,0.421-1.288,1.021-1.695,1.738c-0.634,1.114-0.796,2.408-0.457,3.643
	C-6.202-6.342-5.4-5.313-4.286-4.68c0.24,0.136,0.324,0.441,0.188,0.682C-4.19-3.836-4.359-3.745-4.533-3.745z";
\fill[#1] svg "M-5.88-1.348c-0.083,0-0.168-0.021-0.246-0.065c-1.986-1.125-3.416-2.956-4.025-5.156c-0.609-2.2-0.325-4.505,0.799-6.492
	c0.723-1.274,1.77-2.35,3.025-3.109C-6.09-16.313-5.782-16.236-5.64-16c0.143,0.236,0.067,0.544-0.169,0.687
	c-1.11,0.671-2.034,1.621-2.673,2.747c-0.993,1.753-1.244,3.789-0.706,5.731c0.538,1.943,1.8,3.56,3.554,4.553
	c0.241,0.136,0.325,0.441,0.189,0.682C-5.536-1.439-5.706-1.348-5.88-1.348z";
\fill[#1] svg "M-7.574,1.059c-0.088,0-0.177-0.023-0.257-0.071c-2.627-1.581-4.482-4.089-5.223-7.065c-0.74-2.975-0.277-6.061,1.303-8.689
	c0.958-1.591,2.303-2.94,3.892-3.9c0.237-0.144,0.544-0.067,0.687,0.169c0.143,0.236,0.067,0.544-0.169,0.687
	c-1.45,0.877-2.679,2.108-3.553,3.561c-1.442,2.399-1.865,5.216-1.189,7.932c0.675,2.716,2.369,5.006,4.768,6.449
	c0.236,0.143,0.313,0.45,0.171,0.687C-7.239,0.972-7.405,1.059-7.574,1.059z";
\fill[#1] svg "M4.536-3.745c-0.174,0-0.343-0.091-0.435-0.253C3.964-4.238,4.048-4.543,4.289-4.68c1.113-0.633,1.914-1.662,2.253-2.898
	c0.34-1.235,0.178-2.53-0.456-3.644c-0.404-0.712-0.99-1.313-1.694-1.738c-0.237-0.143-0.313-0.45-0.17-0.686
	c0.142-0.238,0.45-0.314,0.686-0.17c0.851,0.513,1.56,1.239,2.048,2.101c0.766,1.346,0.961,2.91,0.551,4.403
	c-0.41,1.493-1.377,2.737-2.723,3.502C4.705-3.766,4.62-3.745,4.536-3.745z";
\fill[#1] svg "M5.882-1.348c-0.174,0-0.344-0.091-0.436-0.253C5.31-1.841,5.395-2.146,5.635-2.283c3.62-2.05,4.897-6.664,2.848-10.285
	c-0.64-1.127-1.564-2.077-2.674-2.746c-0.236-0.143-0.312-0.45-0.17-0.687c0.143-0.237,0.449-0.313,0.687-0.17
	c1.256,0.758,2.302,1.833,3.027,3.109c2.322,4.101,0.875,9.326-3.224,11.648C6.05-1.369,5.965-1.348,5.882-1.348z";
\fill[#1] svg "M7.576,1.059c-0.17,0-0.335-0.086-0.429-0.242C7.005,0.58,7.082,0.272,7.318,0.13c4.95-2.978,6.555-9.429,3.576-14.381
	c-0.873-1.451-2.101-2.683-3.55-3.561c-0.236-0.143-0.312-0.451-0.168-0.687c0.144-0.236,0.451-0.311,0.687-0.168
	c1.587,0.962,2.932,2.311,3.889,3.9c3.263,5.425,1.505,12.492-3.917,15.753C7.753,1.036,7.664,1.059,7.576,1.059z";
},minimum width=2.2em,minimum height=3em},
Rsu/.default=black}

\tikzset{Server/.style={path picture={
\pgftransformscale{3em/100cm}
\pgftransformyscale{-1}
\fill[#1] (-34.646,-26.568) -- (-34.648,36.998) -- (-5.069,42.52) -- (-5.069,-22.815) -- cycle;
\fill[white] (-18.734,37.15) -- (-21.861,36.579) -- (-21.861,7.092) -- (-18.734,7.43) -- cycle;
\fill[white] (-9.257,5.562) -- (-30.484,2.862) -- (-30.484,-3.595) -- (-9.257,-0.895) -- cycle;
\fill[white] (-9.257,-3.595) -- (-30.484,-6.296) -- (-30.484,-12.754) -- (-9.257,-10.053) -- cycle;
\fill[white] (-9.257,-12.755) -- (-30.484,-15.456) -- (-30.484,-21.914) -- (-9.257,-19.212) -- cycle;
\fill[#1] (-4.187,-24.749) -- (32.57,-39.392) -- (7.907,-42.52) -- (-32.506,-28.34) -- cycle;
\fill[#1] (-3.039,-23.021) -- (-3.039,42.166) -- (34.648,21.965) -- (34.648,-38.036) -- cycle;
},minimum width=3em,minimum height=3em},
Server/.default=black}

\graphicspath{{./}{icons/}}
\usepackage{soul,color}
\usepackage{xcolor}

\usepackage{cite}

\begin{document}
%
\title{Latency and Throughput Optimization in Modern Networks: A Comprehensive Survey}
%
%
%

\author{Amir~Mirzaeinnia,~
        Mehdi~Mirzaeinia,~
        and~Abdelmounaam~Rezgui~
\thanks{A. Mirzaeinia is with the Department
of Computer Science and Engineering, New Mexico Institute of mining and Technology, Socorro, NM, 87801 USA e-mail: amirzaei@cs.nmt.edu
	}
\thanks{A. Rezgui is with Illinois State University.}
\thanks{Manuscript received XXXX; revised XXXX.}}

%
%

\markboth{ready to submit to  IEEE Communications Surveys \& Tutorials  Journal
}%
{Shell \MakeLowercase{\textit{et al.}}: Bare Demo of IEEEtran.cls for IEEE journals}
%



\maketitle


\begin{abstract}
Modern applications are highly sensitive to communication delays and throughput. This paper surveys major attempts on reducing latency and increasing the throughput. These methods are surveyed on  different networks and surrondings such as  wired  networks, 
wireless networks, 
application layer transport control, 
Remote Direct Memory Access, 
and machine learning based transport control, 

\end{abstract}



%
\IEEEpeerreviewmaketitle

\section{Introduction}

Recent applications such as Virtual Reality (VR), autonomous cars or aerial vehicles, and telehealth need high throughput and low latency communication. These applications have to communicate through variety of network and traffic characteristics that are evolved into massive diverse cases. High Bandwidth Product (HBD), Multi paths, Data Centers, and a variety of wireless networks are some of the main domains that their traffic needs to be managed properly. Latency and throughput optimization (L\&T) are even more complex problems to achieve in such a huge environment. Network congestion, rate control, and load balancing are some of the main approaches to imrpove the latency and throughput of the different networks. We have found that, thus far, different networks need customized control mechanism to  achieve optimum performance. However Machine Learning (ML) algorithms (specially deep learning based) are a promising approach to achieve such an ambitious adaptive intelligent mechanism to optimally work in different network environments and serve different traffic characteristics. In 2017 Silver et al. \cite{silver2016mastering,silver2017mastering}  designed a deep learning based computer program to play Go that was able to defeat a world champion challenge. These results show that ML algorithms are able to achieve beyond human capability. Games are the best benchmark problems to test machine learning algorithms to control a sequence of decisions. Network data rate, congestion control and load balancing as main parts contributing on (L\&T) are also a decision sequence making problem that could be resolved through deep learning algorithms. In this paper, we present a comprehensive survey of the network rate and congestion control problem starting from the beginning till recent intelligent approaches.

On one hand every user likes to send and receive their data as quickly as possible. On the other hand the network infrastructure that connects users has limited capacities and these are usually shared among users. There are some technologies that dedicate their resources to some users but they are not very much commonly used. The reason is that although dedicated resources are more secure they  are more expensive to implement. Sharing a  physical channel among multiple transmitters needs a technique to control their rate in proper time. The very first congestion network collapse was observed and reported by Van Jacobson in 1986. This caused about a thousand time rate reduction from 32kbps to 40bps \cite{Jacobson} which is about a thousand times rate reduction. Since then very different variations of the Transport Control Protocol (TCP) have been designed and developed.

There are some main capabilities required to achieve. First, handling the congestion may cause the delay of a certain flow which is not desired for the network users. Resource maximum utilization to serve  more users is also desired. Therefore increasing the number of users and their transmission rate is desired and this increasing makes it more challening to optimize user's latency and throughput. Transport control is an approach to improve the (L\&T) of the networks. There are two main class of

TCP is known as connection oriented transmission which means senders wait to receive an acknowledgement signal packet. To continue transmission the sender needs to receive acknowledgment of the previous packet in a certain timeout period. Otherwise  the previous packet which is not acknowledged yet would be retransmitted. Packet by packet data transmission and waiting for its acknowledgment wastes time and network resources. Therefore the idea of sliding windows evolved to control the transmission rate more efficiently while it can also check the acknowledgment of the transmitted packet. Additionally there are circular sequence numbers to keep track of the transmitted packets.

Generally, there are different classes of approaches proposed to control the rate in networks. Rate based \cite{Xu2005} ( streaming media protocols) and window based are the two most common groups of congestion control mechanisms. Different traffic packet scheduling is another way to address the congestion problem in the networks which is covered in more detail  in  \cref{sec_scheduling}.

\begin{figure}
	
	\centering
	\resizebox{.5\textwidth}{!}{
	\begin{tikzpicture}
	\path (0,0)  node(ag) []   {}
	(6,4) node(a) [rounded rectangle, draw, fill=green!30]   {\LARGE CC Feedback signals} 
	(6,3) node(aux1) [ ]   {} 
			
	(1.5,2) node(b) [rounded rectangle, minimum width=2cm,draw,fill=green!60]  {\large LOSS} 
	(4.5,2) node(c) [rounded rectangle, minimum width=2cm,draw,fill=green!60]  {\large RTT} 
	(7.5,2) node(d) [rounded rectangle, minimum width=2cm,draw,fill=green!60]  {\large ECN} 
	(10.5,2) node(e) [rounded rectangle, minimum width=2cm,draw,fill=green!60]  {\large RTT Gradient}

	(2.2,1)      node(b1) [rectangle, minimum width=2.2cm,draw,fill=pink!60] {TCP Tahow}
	(2.2,.5)    node(b2) [rectangle, minimum width=2.2cm,draw,fill=pink!60] {TCP Reno} 
	(2.2,0)     node(b3) [rectangle, minimum width=2.2cm,draw,fill=pink!60] {SACK}
	(2.2,-.5)  node(b4) [rectangle, minimum width=2.2cm,draw,fill=pink!60] {BIC}

	(5.2,1)      node(c1) [rectangle, minimum width=2.2cm,draw,fill=pink!60] {TCP Vegas}
	(5.2,.5)    node(c2) [rectangle, minimum width=2.2cm,draw,fill=pink!60] {Fast TCP} 
	(5.2,0)     node(c3) [rectangle, minimum width=2.2cm,draw,fill=pink!60] {Westwood}
	(5.2,-.5)  node(c4) [rectangle, minimum width=2.2cm,draw,fill=pink!60] {Jersey}
	
	(8.2,1)      node(d1) [rectangle, minimum width=2.2cm,draw,fill=pink!60] {DCTCP \cite{Alizadeh2011}}
	(8.2,.5)    node(d2) [rectangle, minimum width=2.2cm,draw,fill=pink!60] {DCCP \cite{kohler2006datagram}} 
	(8.2,0)     node(d3) [rectangle, minimum width=2.2cm,draw,fill=pink!60] {XCP \cite{Katabi2002}}
	
	(12,1)  node(e1) [rectangle, minimum width=3cm,draw,fill=pink!60] {Timely \cite{mittal2015timely}}
	(12,0.5)      node(e2) [rectangle, minimum width=3cm,draw,fill=pink!60] {Others \cite{hayes2011revisiting,armitage2013using}}
	
	;

	\draw [ line width=2]  (a.south) -| (aux1.north);
	\draw [->, line width=2]  (aux1.north) -| (b.north);
	\draw [->, line width=2]  (aux1.north) -| (c.north);
	\draw [->, line width=2]  (aux1.north) -| (d.north);
	\draw [->, line width=2]  (aux1.north) -| (e.north);
				
	\foreach \value in {1,...,4}
	\draw[->] (b.200) |- (b\value.west);
	
	\foreach \value in {1,...,4}
	\draw[->] (c.200) |- (c\value.west);
	
	\foreach \value in {1,...,3}
	\draw[->] (d.200) |- (d\value.west);
	
	\foreach \value in {1,...,2}
	\draw[->] (e.205) |- (e\value.west);

	\end{tikzpicture}
}
	\caption{feedback signals.} 
	\label{fig_cc_signals}
\end{figure}
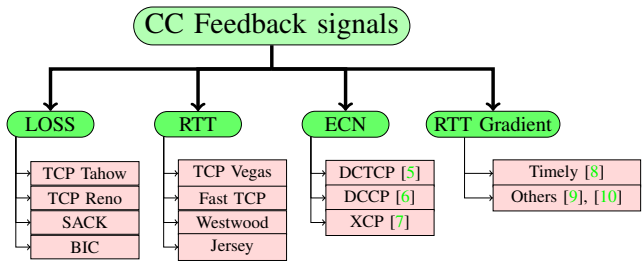


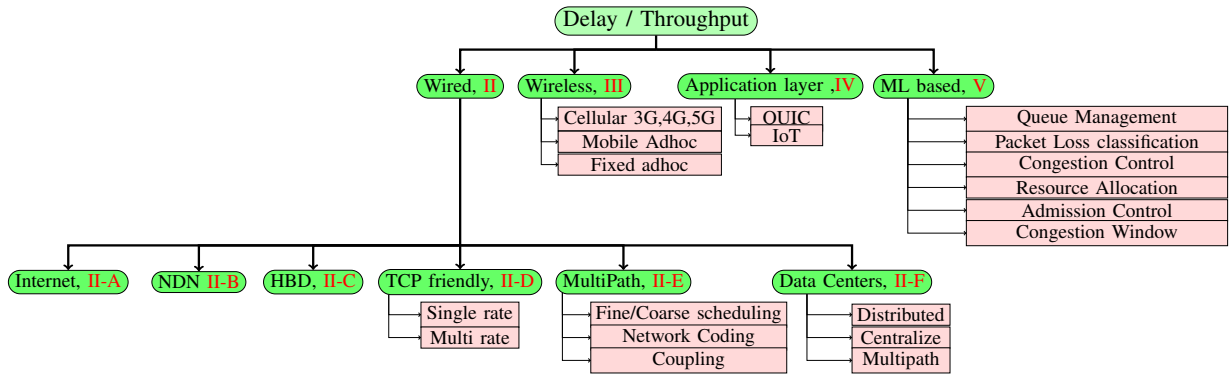
\begin{figure*}
	
	\centering
	\resizebox{.9\textwidth}{!}{
		\begin{tikzpicture}
		\path (0,0)  node(ag) []   {}
		(8,4) node(a) [ rounded rectangle, draw, fill=green!30]   {\huge Delay / Throughput} 
		(8,3) node(aux1) [ ]   {} 
		
		(2,2) node(b) [rounded rectangle, minimum width=2cm,draw,fill=green!60]  {\LARGE Wired, \ref{sec_wired}} 
		(2,-3) node(aux2) [ ]   {} 
		(5.5,2) node(c) [rounded rectangle, minimum width=2cm,draw,fill=green!60]  {\LARGE Wireless, \ref{sec_wireless}} 
		(11.5,2) node(d) [rounded rectangle, minimum width=2cm,draw,fill=green!60]  {\LARGE Application layer ,\ref{sec_app_cc}} 
		(16.5,2) node(e) [rounded rectangle, minimum width=2cm,draw,fill=green!60]  {\LARGE ML based, \ref{sec_ML}}

	(-10,-4) node(b2) [rounded rectangle, minimum width=2cm,draw,fill=green!60]  {\LARGE Internet, \ref{sec:internet}} 
	(-6,-4)  node(b1) [rounded rectangle, minimum width=2.2cm,draw,fill=green!60] {\LARGE NDN \ref{sec:Namednetwork}} 
	(-2.5,-4) node(b3) [rounded rectangle, minimum width=2cm,draw,fill=green!60]  { \LARGE HBD, \ref{sec_nigh_BDP}} 
	(2,-4) node(b4) [rounded rectangle, minimum width=2cm,draw,fill=green!60]  {\LARGE TCP friendly, \ref{tcp_friendly}} 
	(7,-4) node(b5) [rounded rectangle, minimum width=2cm,draw,fill=green!60]  {\LARGE MultiPath, \ref{sec_multipath}} 
	(14,-4) node(b6) [rounded rectangle, minimum width=2cm,draw,fill=green!60]  {\LARGE Data Centers, \ref{sec_DC}}


		(2.3,-5)   node(b31) [rectangle, minimum width=3cm,draw,fill=pink!60] {\LARGE Single rate} 
		(2.3,-5.7)    node(b32) [rectangle, minimum width=3cm,draw,fill=pink!60] {\LARGE Multi rate}

		(9,-5)    node(b41) [rectangle, minimum width=6cm,draw,fill=pink!60] {\LARGE Fine/Coarse scheduling} 
		(9,-5.7)    node(b42) [rectangle, minimum width=6cm,draw,fill=pink!60] {\LARGE Network Coding}
		(9,-6.4)    node(b43) [rectangle, minimum width=6cm,draw,fill=pink!60] {\LARGE Coupling} 
		
		(15.5,-5)    node(b51) [rectangle, minimum width=3cm,draw,fill=pink!60] {\LARGE Distributed} 
		(15.5,-5.7)    node(b52) [rectangle, minimum width=3cm,draw,fill=pink!60] {\LARGE Centralize} 
		(15.5,-6.4)    node(b53) [rectangle, minimum width=3cm,draw,fill=pink!60] {\LARGE Multipath } 
		
		(7.5,1)      node(c1) [rectangle, minimum width=5cm,draw,fill=pink!60] {\LARGE Cellular 3G,4G,5G}
		(7.5,.3)     node(c3) [rectangle, minimum width=5cm,draw,fill=pink!60] {\LARGE Mobile Adhoc}
		(7.5,-.4)    node(c2) [rectangle, minimum width=5cm,draw,fill=pink!60] {\LARGE Fixed adhoc} 
		
		(12,1)      node(d1) [rectangle, minimum width=2.2cm,draw,fill=pink!60] {\LARGE QUIC}
		(12,.5)    node(d2) [rectangle, minimum width=2.2cm,draw,fill=pink!60] {\LARGE IoT}

		(21.5,1)    node(e1) [rectangle, minimum width=8cm,draw,fill=pink!60] {\LARGE Queue Management} 
		(21.5,.3)    node(e2) [rectangle, minimum width=8cm,draw,fill=pink!60] {\LARGE Packet Loss classification} 		
		(21.5,-.4)    node(e4) [rectangle, minimum width=8cm,draw,fill=pink!60] {\LARGE Congestion Control} 				
		(21.5,-1.1)    node(e6) [rectangle, minimum width=8cm,draw,fill=pink!60] {\LARGE Resource Allocation} 
		(21.5,-1.8)    node(e5) [rectangle, minimum width=8cm,draw,fill=pink!60] {\LARGE Admission Control} 					
		(21.5,-2.5)    node(e3) [rectangle, minimum width=8cm,draw,fill=pink!60] {\LARGE Congestion Window} 		
	

		; 
		
		\draw [->, line width=2]  (aux1.north) -| (b.north);
		\draw [->, line width=2]  (aux1.north) -| (c.north);
		\draw [->, line width=2]  (aux1.north) -| (d.north);
		\draw [->, line width=2]  (aux1.north) -| (e.north);
		\draw [ line width=2]  (a.south) --(aux1.north);

		\draw [ line width=2]  (b.south) --(aux2.north);
		\draw [->, line width=2]  (aux2.north) -| (b1.north);
		\draw [->, line width=2]  (aux2.north) -| (b2.north);
		\draw [->, line width=2]  (aux2.north) -| (b3.north);
		\draw [->, line width=2]  (aux2.north) -| (b4.north);
		\draw [->, line width=2]  (aux2.north) -| (b5.north);		
		\draw [->, line width=2]  (aux2.north) -| (b6.north);		
		
		\foreach \value in {1,...,3}
		\draw[->] (c.200) |- (c\value.west);
		
		\foreach \value in {1,...,2}
		\draw[->] (d.200) |- (d\value.west);
		
		\foreach \value in {1,...,6}
		\draw[->] (e.205) |- (e\value.west);
		
			\foreach \value in {1,...,3}
		\draw[->] (b6.195) |- (b5\value.west);
		
			\foreach \value in {1,...,2}
		\draw[->] (b4.190) |- (b3\value.west);
		
			\foreach \value in {1,...,3}
		\draw[->] (b5.190) |- (b4\value.west);
		

		\end{tikzpicture}
	}
	\caption{Paper Structure.} 
	\label{fig:paper_structure}
\end{figure*}

Congestion Control mechanisms could be also classified by incremental modification. Only sender needs modification; sender and receiver need modification; middle nodes (like switches or routers) need modification; sender, receiver and middle nodes need modification. Another way to classify the CC mechanisms could be based on performance metrics that they tackle. High bandwidth-delay product networks, lossy links, fairness, advantage to short flows, and  variable-rate links are some of the most commonly tried mechanisms. Congestion control mechanisms could be also classified by fairness criteria such as  Max-min fairness and proportionally fair. Authors in \cite{Abbas2016} surveyed the fairness-driven queue management mechanisms.

When a CC protocol is about to be designed, there are some key points that need to be taken into account. 1- Competitive flows have no information from each other. 2-The number of competitive flows are not known to flow sources. 3-Flows sources are not aware of available resources and bandwidth.

Network and traffic transmitters need to react against congestions they can sense. Almost all network devices are equipped with some input, output or input/output buffers \cite{firoozshahian2007efficient,appenzeller2004sizing,lin2005throughput,mckeown1999achieving_VOQ} to manage the congestion to some extent. However network excessive congestion causes the network devices buffer to overflow. Drop tail is the first method that is performed in routers/switches to control the incoming of excessive traffic. Drop tail causes a situation called flow synchronization. Flow synchronization reduces the lower bandwidth utilization. Different Active Queue Management have been proposed to manage network buffering. AQM schemes are surveyed in \cite{adams2012active}. Random Early Detection (RED) is designed to overcome the flow synchronization issue, and Weighted RED (WRED) is  developed to provide fairness to responsive TCP and non-responsive UDP traffic at a buffer of a communications network \cite{olesinski2009fair}. Traditional congestion control protocols work based on additive increase and multiplicative decrease. Rate decrease always happened to halve the transmission rate. Besides,  current congestion control protocols can be classified into two types: 1- the global end-to-end-based congestion control \cite{bhatti2008revisiting,ho2007ward,miras2008fairness,hasegawa2001survey,Head2010,Goyal,Kunniyur2001} and 2- the local link-based congestion control \cite{Sivaraman,Mittala,kong_improving_2018,liu2008tcp,NEAL2016,Widmer,Amherst2000,JiaZhao2015,Reddy2017}. Afanasyev et al.  \cite{Afanasyev2010} published an example of end to end control mechanisms.

\tikzstyle{l} = [draw, -latex',thick]
\begin{figure}
	\centering

\begin{tikzpicture}

\path (0,0)  node(ag) []    {}
 (4,3) node(a) [rounded rectangle,draw,fill=green!60]               {Performance Metrics} 
 (4,2.3) node(aux1) []               {} 
 
(1,-0) node(b) [rectangle,rotate=90,minimum width=4cm,draw,fill=pink!60]               {Efficiency} 
(2,-0) node(c) [rectangle,rotate=90,minimum width=4cm,draw,fill=pink!60]               {Fairness} 
(3,-0) node(d) [rectangle,rotate=90,minimum width=4cm,draw,fill=pink!60]               {Smoothness}

(4,0)  node(e) [rectangle,rotate=90,minimum width=4cm,draw,fill=pink!60]  {Responsiveness} 
(5,0)  node(f) [rectangle,rotate=90,minimum width=4cm,draw,fill=pink!60]  {Aggressiveness} 
(6,-0) node(g) [rectangle,rotate=90,minimum width=4cm,draw,fill=pink!60]               {Convergence speed} 
(7,0)  node(h) [rectangle,rotate=90,minimum width=4cm,draw,fill=pink!60] {Flow Completion Time}
;

	\draw [ line width=1]  (a.south) --(aux1.north);
\draw [->, line width=1]  (aux1.north) -| (b.east);
\draw [->, line width=1]  (aux1.north) -| (c.east);
\draw [->, line width=1]  (aux1.north) -| (d.east);
\draw [->, line width=1]  (aux1.north) -| (e.east);
\draw [->, line width=1]  (aux1.north) -| (f.east);		
\draw [->, line width=1]  (aux1.north) -| (g.east);		
\draw [->, line width=1]  (aux1.north) -| (h.east);

\end{tikzpicture}

\caption{Performance Metrics.} 
\label{fig_Performance_Metrics}
\end{figure}
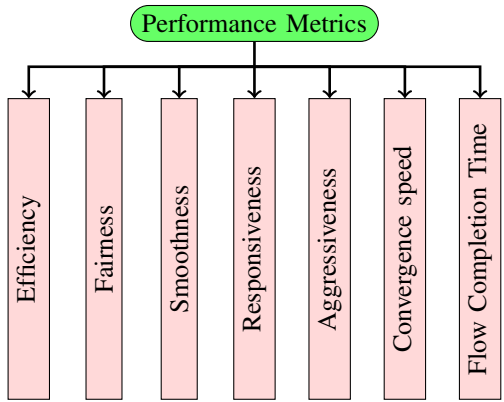

Rest of this section covers the congestion detection,  notification ,  control, and the performance metrics that are mostly used in CC mechanisms.
\subsection{Latency and Throughput (L\&T)}
\subsubsection{Congestion Detection}
Congestion detection can be through many different network statistics such as packet loss, queue length, queue length and channel load, channel busyness ratio and throughput measurement, packet service time, packet service time and queue length, ratio of packet service time, and packet inter-time (scheduling time). Figure \ref{fig_cc_signals}. shows the main signals to detect the network congestion.

Drop/loss based. TCP flows wait for the packet to receive acknowledgement to continue with other packets to transmit. Time out is also used to control the time each transmitter has to wait until receiving the acknowledgment. If  transmitter does not receive acknowledgement by end of the timeout period then it is assumed that either transmitted packet or its acknowledgment is lost (either dropped or failed). 
Packet loss is sensed after a time out or receiving a NAK packet. For example Additive Increase Multiplicative Decrease (AIMD) based TCP-Tahoe is slow to start when a drop happens, TCP-NewReno and TCP-selective ACK (SACK) \cite{mathis1996tcp,sikdar2003analytic} are a more conservative slow start when a  three duplicated acknowledge packet is sensed, AIMD-FC \cite{lahanas2002additive}, Binomial Mechanisms \cite{Bansal}, HIGHSPEED-TCP (for high BDP)\cite{floyd2003highspeed}, BIC-TCP\cite{li2007experimental}. As another step, General AIMD Congestion Control  GAIMD is  proposed to modify the window size increase/decrease rate \cite{Yang}.

RTT Based. RTT based mechanisms measure the round trip time for each packet to find the network congestion level. Different congestion levels cause different round trip times. Therefore the transmitter can react based on the measured time to control the network congestion. TCP vegas \cite{brakmo1994tcp}, fast TCP\cite{JinDavidWeiStevenH2004},  TCP-WESTWOOD\cite{mascolo2001tcp},   TCP-JERSEY\cite{xu2004tcp} are some of the known mechanisms that relys on network RTT.  RTT based mechanisms usually suffer from unfairness because the flows with shorter RTT are able to attain more shared bandwidth when they compete with flows with longer RTT. These mechanisms are hard to develop in networks where RTT is in the microsecond range such as Data Center networks.

Explicit Congestion Notification (ECN) Based. In ECN based mechanisms  traffic packets are marked by intermediary nodes in their way when switch buffers are build up to a certain threshold.  \cite{Alizadeh2011,Alizadeh2010}. Therefore transmitters start to reduce the transmission rate. ECN flag bit will be set in middle way routers if they encounter a certain level of congestion.

RTT Gradient, recently gradient of the RTTs are proposed as more informative feedback from the networks \cite{hayes2011revisiting,armitage2013using,mittal2015timely}. Linear/non-linear combinations of above mentioned signals are also considered on recent approaches \cite{Donga2015} as utility functions that describe an objective like high throughput and low loss rate.

\subsubsection{Congestion Notification}

Re-transmission Time Out (RTO) and NACK are the two main techniques for traffic transmitters to be notified. RTO is the time that transmitters wait to receive the acknowledgment. Transmitters have to resend its traffic after the RTO time period. A fixed RTO would work in wired networks, however to improve the performance of the wireless network, adaptive RTO is needed because of user mobility. In the case of the NACK mechanism, the receiver sends back a Not acknowledged packet (NACK) if it receives a packet which is out of order. For example, if the receiver received packet number one and three then it NACK back for packet number two.

\subsubsection{Congestion Control (CC)}

A wide variety of network congestion control have been proposed to adapt network and traffic characteristics. In this paper we cover different congestion controls designed for internet, data centers, and wireless networks. Congestion control mechanisms are mostly designed as reactive. However some proactive mechanisms are also published such as \cite{Mozo2018}. This paper proposes a control mechanism that explicitly computes the optimal sessions’ sending rates independently of congestion signals (i.e., proactive mechanisms). Then predicts the proper rate in order to avoid the congestion problems.

\subsubsection{Performance Metrics}

Various metrics have been introduced to compare the performance of different mechanisms. These metrics are listed in Figure \ref{fig_Performance_Metrics}. Some of these metrics help to compare the quality of service that is served to the users and some others help to observe the network resource utilization. Network resource utilization and flow completion time are the two metrics that are most helpful for the network provider and end users respectively.

\paragraph{Efficiency}
According to \cite{mamatas2007approaches} ,”Efficiency is the average flow throughput per step (or per RTT), when the system is in equilibrium”. Equilibrium is the state that the network is stable or competitive flows achieve a balanced and stable resources sharing point.

\paragraph{Fairness}
Best effort nature of the packet based networks is one of the main reasons behind unfairness in network traffic. Unfairness and bandwidth starvation may happen due to having both TCP based and (User Datagram Protocol) UDP based traffics on the same network resources. UDP based flows are bandwidth hungry, prefer a stable data rate, and transmit at a fixed rate. Therefore TCP/UDP flows transmission in the same resources may cause unfair resource allocation. Application layer mechanisms control the congestion in UDP based flows ( \cref{sec_app_cc}).

Besides different TCP protocols may behave differently, and therefore inter-protocol unfairness occurs between different congestion control mechanisms \cite{bhatti2008revisiting,ho2007ward,miras2008fairness,hasegawa2001survey,Goyal}. For example TCP Vegas suffers from bandwidth sharing when it is passed through the same resources with TCP Reno. The reason is that TCP Vegas uses more conservative mechanisms like different slow start and  fast retransmission.

Furthermore unfairness may also happen due to a different Round Trip Time (RTT). This phenomenon is more obvious when RTT based congestion control is applied in the Data Center network where intra-rack traffics are competing with inter-rack traffics ( \cref{sec_DC}).

\paragraph{Smoothness}
Smoothness measures the variation in the transmission rate of a connection using the protocol. Smoothness is important in steady state

\paragraph{Responsiveness}
Responsiveness is another metric used to show the speed needed to reach an equilibrium \cite{zhang2002interrelation}. In other words it measures how fast a connection reacts to increased congestion by decreasing its window size. It is desirable that the connection reduces its transmission rate to its fair share promptly.

\paragraph{Aggressiveness}
Aggressiveness measures how fast the connection probes extra bandwidth by opening up its window. It is particularly desirable that the connection acquires the extra available resources quickly when available bandwidth increases suddenly.

\paragraph{Convergence speed}
Flow start/stop are the two prevalent events in networks. Therefore other flow rate adaptation speed is one of the key factors to measure the performance of control protocols. Delayed or slow response would affect user flow completion time as well as network resource utilizations. Convergence measures how fast competing flows converge to their fair share of bandwidth. In other words, convergence speed is related to the aggressiveness and responsiveness indices. More aggressive and responsive protocols usually converge faster.

\paragraph{Flow Completion Time(FCT)}
FCT is a metric that would be more desired for the end user rather than network service provider \cite{dukkipati2006flow}. Network engineers have been working on other metrics such as bandwidth improvement and efficiency to improve their network performance while end users might not even sense these improvements.

In this paper we reviewed papers published on various transport control protocols designed to improve the performance for different network and traffic characteristics (Figure \ref{fig:paper_structure}).

In  \cref{sec_wired} latency and throughput optimization (L\&T) in wired networks are reviewed. Internet in \cref{sec:internet}, and Named Data Networks in \cref{sec:Namednetwork}, high bandwidth delay product communications in \cref{sec_nigh_BDP} , TCP friendly in \cref{tcp_friendly}, Multi path transport control  in \cref{sec_multipath} protocols,  are considered as wired network control protocols.

In  \cref{sec_DC} Data Center networks, traffic characteristics and (L\&T) mechanisms are studied. This section also covers DC network topologies including switched centric (such as Fattree and Leaf/Spine) and server centric networks (such as BCube and Dcell). DC traffic characteristics are also discussed in  this section that covers incast problems in addition to delay sensitive mice and throughput sensitive elephant flows proportion. Remote Direct Memory Access  (RDMA) is also discussed as the recent advancement on DC networks. \\


In  \cref{sec_wireless} (L\&T) in wireless networks are surveyed.  \cref{sec_cellular} covers  cellulars (3G, 4G, 5G), \cref{sec_sensor_Network} fixed ad hoc networks, \cref{sec_Manet_Vanet} Mobile Ad hoc Networks (MANET), \cref{sec_vanet} Vehicular  Ad hoc Networks (VANET), \cref{sec_fanet} Flying Ad hoc Networks (FANET) are covered.\\

Application layer flow  control mechanism to optimize the (L\&T) are presented in \cref{sec_app_cc},  including CoAP  \cref{sec_IoT} that is designed for Internet of Things (IoT) and Quic \cref{sec_quic} that is designed and implemented by Google to reduce the web access latency.\\

In  \cref{sec_ML} Machine Learning based approaches are discussed to optimize (L\&T) performance. ML algorithms as predictive, descriptive and controllers will be introduced \cref{sec_MLintro}. The rest of the section covers different ML algorithms applied on queue management \ref{sec_ML_QManagement},  packet loss classification\ref{sec_ML_LossClasification}, congestion conrol \ref{sec_ML_CC}, resource allocation \ref{sec_ML_resourceAllocation}, admission control \ref{sec_ML_Admission_control}, congestion window control \ref{sec_ML_Congestion Window}.

As the final part of this paper,  lessons learned and research opportunities are presented in   \cref{sec_conclusion} .

\section{Latency and Throughput Optimization (L\&T) in Wired  Networks}\label{sec_wired}

Kunniyur. et, al. in \cite{Kunniyur2001} proposed Adaptive Virtual Queue (AVQ) and investigate it's stability in the presence of feedback delays, it's  ability to maintain small queue lengths and  robustness in the presence of extremely short flows (the so-called  mice flows). Authors in \cite{Sivaraman,Mittala} proposed two other approaches in network congestion control.

\subsection{Internet}\label{sec:internet}

Modern networks are massively diverse in terms of conditions. For example, RTTs of competing flows may vary by more than 1000X; link bandwidths are from Kbps to Gbps; real time vs non real time traffic transmission; multipath traffic transmission possibility. Performance-oriented Congestion Control (PCC)\cite{Donga2015} sends at a rate r for a short period of time, and observes the results SACKs indicating delivery, loss, and latency of each packet. It aggregates these packet-level events into a utility function that describes an objective like “high throughput and low loss rate”. The result is a single numerical performance utility. It runs a micro experiment to control the rate, and continuous multiple micro experiments help to achieve empirically consistent high performance. PCC only needs  a sender modification and they are experimented on with different network characteristics such as in the wild Internet, satellite link, unreliable lossy links, unequal RTT competing sender, rapidly changing networks.

Sundaresan et al. in \cite{Sundaresan} developed an Internet path measurement technique to distinguish congestion experienced when a flow self-induced congestion in the path from when a flow is affected by an already congested path. This technique can be applied to speed tests when the user is affected by congestion either in the last mile or in an interconnect link. This difference is important because in the latter case the user is constrained by their service plan (i.e., what they are paying for), and in the former case they are constrained by forces outside of their control.These days CUBIC is known as a default LINUX TCP algorithm.  CUBIC modifies the linear window growth to be a cubic function in order to improve the scalability of TCP over fast and long distance networks. Window size in CUBIC slowly increases when it is close to the saturation point  \cite{Haa2008}. Authors in \cite{Sundaresan2017,NEAL2016}, proposed other control mechanisms for internet services.

Nimbus\cite{goyal2018elasticity} introduces a method to detect elasticity of the cross competing traffic. In this method the sender modulates its rate with sinusoidal pulses to create small traffic fluctuations at the bottleneck at a specific frequency (e.g., 5 Hz). Then Nimbus concurrently estimates the rate of cross traffic based on its own send and receive rates. Nimbus monitors frequency response (FFT) of the cross traffic to determine if the cross traffic’s rate oscillates at the same frequency. If cross traffic is inelastic then the sender can control queueing delays while achieving high throughput. However, in case of elastic traffic, the sender may lose throughput if it attempts to control packet delay.

 Low-Latency Low-Loss Scalable-Throughput (L4S) \cite{Briscoe2019L4s} has been proposed as a new internet service . Nadas et al. \cite{nadas2020congestion} showed that L4S schedulers are not able to handle high levels of Internet heterogeneity. They propose a virtual dual queue named Core Stateless Active Queue Management (CSAQM) concept to maintain separate queues for L4S and classic traffic. Therefore a coupled packet dropping/marking mechanism in CSAQM improves the fairness among different flows.

\subsection{Named Data Networks} \label{sec:Namednetwork}
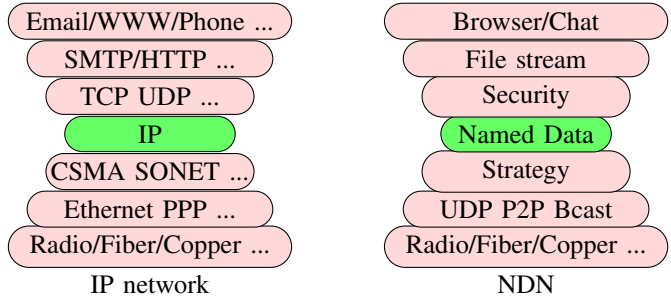
\begin{figure}[h]
	\centering
	
	\begin{tikzpicture}
	
	\path (0,0)  node(ag) []    {}
	(6,-.5) node(b) [rounded rectangle,minimum width=4cm]               {NDN} 
	(6,-0) node(b) [rounded rectangle,minimum width=4cm,draw,fill=pink!60]               {Radio/Fiber/Copper ...} 
	(6,.5) node(c) [rounded rectangle,minimum width=3.5cm,draw,fill=pink!60]               {UDP P2P Bcast} 
	(6,1) node(e) [rounded rectangle,minimum width=3cm,draw,fill=pink!60]               {Strategy}

	(6,1.5)  node(f) [rounded rectangle,minimum width=2.5cm,draw,fill=green!60]  {Named Data} 
	(6,2)  node(f) [rounded rectangle,minimum width=3cm,draw,fill=pink!60]  {Security} 
	(6,2.5) node(d) [rounded rectangle,minimum width=3.5cm,draw,fill=pink!60] {File stream} 
	(6,3)  node(g) [rounded rectangle,minimum width=4cm,draw,fill=pink!60] {Browser/Chat}

	(1,-.5) node(b) [rounded rectangle,minimum width=4cm,]               {IP network} 
	(1,-0) node(b) [rounded rectangle,minimum width=4cm,draw,fill=pink!60]               {Radio/Fiber/Copper ...} 
	(1,.5) node(c) [rounded rectangle,minimum width=3.5cm,draw,fill=pink!60]               {Ethernet PPP ...} 
	(1,1) node(e) [rounded rectangle,minimum width=3cm,draw,fill=pink!60]               {CSMA SONET ...}

	(1,1.5)  node(f) [rounded rectangle,minimum width=2.5cm,draw,fill=green!60]  {IP} 
	(1,2)  node(f) [rounded rectangle,minimum width=3cm,draw,fill=pink!60]  {TCP UDP ...} 
	(1,2.5) node(d) [rounded rectangle,minimum width=3.5cm,draw,fill=pink!60] {SMTP/HTTP ...} 
	(1,3)  node(g) [rounded rectangle,minimum width=4cm,draw,fill=pink!60] {Email/WWW/Phone ...}
	;

	\end{tikzpicture}
	
	\caption{Host centric vs Data centric addressing .} 
	\label{fig:NDN}
\end{figure}

Named Data Networking (NDN) is an emerging internet architecture that is evolving from push mode to pull mode (Figure \ref{fig:NDN}). NDN emphasizes content by making it directly addressable and routable instead of IP addresses.  Authors in \cite{Soniya2015,Saxena2016,Mathematics2017} surveyed the NDN networks. Compared with the traditional TCP/IP networking, the transport in NDN has its specific characteristics such as receiver-driven (pull based),  in-path caching, hop-by-hop, one-Interest-one-data,  multi-Source, and multi-Path.  Hence NDN congestion control mechanisms are also different from classical networks.

\tikzstyle{l} = [draw, -latex',thick]

ICP \cite{Carofiglio2012} is an interest control protocol for content-centric networking. ICP realizes a window-based Interest flow control and achieves high efficiency and fairness under proper parameters setting. Considering the NDN as receiver-driven transport control, the old implicit congestion signals would not be helpful. Ren et al. \cite{Ren2016} proposed a NDN congestion control algorithm based on explicit congestion feedback - ECP (Explicit Control Protocol). It monitors the length of the outgoing interest queue at the intermediate nodes to detect the network congestion status proactively. Then it sends explicit feedback to the receiver to adjust the sending rate of Interests to control the returning rate of data correspondingly. Practical congestion control (PCON) \cite{Schneider2016a}  scheme is also designed to exploit multiple sources and  multiple paths capability of the NDNs. PCON detects congestion based on the CoDel \cite{Nichols2012} AQM (by measuring packet queuing time). Then it explicitly marks certain packets to signals back towards consumers . Therefore downstream routers can divert traffic to alternative paths and consumers can reduce their Interest sending rates. Lei et al. \cite{Lei2015}  explores rate control protocol (RCP), which is a unique hop-by-hop feature in NDN routing, and the one-to-one relationship between interest requests and data packets. They designed  a rate-based, explicit and hop-by-hop congestion control algorithm named NDN Hop-by-Hop RCP (NHBH-RCP).    Content popularity prediction is a CC approach  that is proposed in \cite{Park2014}. Authors in \cite{Ren2016a,Chen2016} surveyed transport mechanisms in NDN networks.


In some references, NDN networks are also called Information-Centric Networking (ICN) \cite{xylomenos2013survey} or Content Centric Networks (CCN).  Authors in \cite{carofiglio2013multipath,janaszka2012popularity,carofiglio2016optimal,conti2020road,carofiglio2019enabling,wang2019design,zhang2019concurrent} studied multipath CC for ICN. Mobile ICN \cite{fang2018survey} are other research trends to enhance network flexibility and mobility management. NDN also benefit from in-network caching \cite{jacobson2009networking}. Authors in \cite{zhang2014transport,tariq2019forwarding,liu2019accp,schneider2019multipath} studied transport protocols in these networks. 


\subsection{High BDP Product Network}\label{sec_nigh_BDP}
Increasing the link bandwidth (fiber optics)  and  delay (inter continent links) introduces other network characteristics that need to be considered when designing new transport control protocols. Pipeline capacity or packet on the fly are the two terms to define the networks that have their high bandwidth and delay multiplications. The efficiency problem of these networks is because traffic on the fly is more than the TCP window size. Therefore traffic senders have to stay idle for some time to receive their acknowledgements. Hence other research trends started and following methods were published.

In 2000, Floyd in \cite{floyd2003highspeed}, proposed TCP's congestion control mechanism to be used in TCP connections with large congestion windows . Later on Cern in \cite{Cern2002}, introduced Scalable TCP (STCP) to improve bulk transfer performance in high speed wide area networks. In 2002, Ketabi et al. \cite{Katabi2002} introduced eXplicit Control Protocol (XCP) which is a general form of the Explicit Congestion Notification proposal (ECN). They also introduce the new concept of decoupling utilization from fairness control. In 2004, Xu et al. introduced the Binary Increase Congestion control (BIC) and considered RTT fairness in  addition to bandwidth scalability. They utilized two window size control policies called additive increase and binary search increase \cite{xu2004binary_BIC}. In 2008, VCP published \cite{YongXia2008} to optimize the XCP by utilizing two ECN bits for network congestion feedback.Ha et al. \cite{Haa2008} modifies the linear window growth function of  TCP  to be a cubic function. This modification improves the scalability of TCP over fast and long distance networks. As cubic function behaves, CUBIC increases the window size aggressively when the window is far from the saturation point, and slowly when it is close to the saturation point. These days, CUBIC is still the default Linux transport protocol. Another protocol is published in \cite{JinDavidWeiStevenH2004} and a comparative study of high-speed Linux TCP variants over high-BDP networks is presented in \cite{Alrshah2014}.

\subsection{TCP-Friendly} \label{tcp_friendly}

Real time applications such as audio/video need minimum delay in data transmission, otherwise they lose their quality and the interest of their users consequently. Additionally these types of traffic can tolerate a little bit of traffic loss. For example losing one packet out of an image frame of a video does not affect the quality of that video very much. Therefore real time applications tend to not deploy the TCP transport protocol. Instead they used to employ more aggressive data transmission protocols such as User Datagram Protocol (UDP). UDP protocol does not provide packet received acknowledgement and is considered a connectionless transmission protocol. Growing too many real time applications these days may cause less chance of TCP based flows. This problem may cause a network collapse by itself, therefore it has been the main reason to develop the TCP-Friendly rate Control protocol (TFRC) \cite{Widmer,Amherst2000}. Considering the real time traffic that can be unicast and multicast traffic, TFRC needs to be designed to cover both two types of traffic. TFRC still suffers from packet loss and delay due to long distance heavy traffic and network fluctuations. Authors in \cite{Reddy2017} introduce a number of key concerns like enhanced Round Trip Time (RTT) and Retransmission Time Out (RTO) calculations, enhanced Average Loss Interval (ALI) methods, and improved Time to Live (TTL). These features are applied to TFRC to boost the performance of TFRC over wired networks. McQuistin et al. \cite{Mcquistin2015} also showed how use of ECN affects the reach- ability of UDP servers. Authors in \cite{Head2010} surveyed TCP friendly protocols. Jin et al. \cite{Jin2001a} propose Square-Increase/Multiplicative-Decrease (SIMD) that is a window-based congestion control algorithm. SIMD uses history information in its control rules. It increases in window size in proportion to the square of the time elapsed since the detection of the last loss event and It uses a multiplicative decrease mechanism.

\subsubsection{Unicast TCP-Friendly}
TCP-friendly rate control (TFRC) uses mathematical models to mimic the TCP equivalent. Equation \ref{eu_tcp_friendly} is the simplest model proposed for the TFRC\cite{Widmer,Amherst2000}.

\begin{equation} \label{eu_tcp_friendly}
T(t_{RTT},s,p) = \frac{c.s}{t_{RTT}.\sqrt{p}}
\end{equation}

Where c is a constant (commonly approximated as $1.5\sqrt{2/3}$), s is the segment size, p is the packet loss rate, and $t_{RTT}$ is round trip time. TFRC uses the windowing mechanism  to control the rate. According to the formula increasing the packet loss rate and also round trip time decreases the transmission rate. Note that there is no packet retransmission for the lost packet in TFRC. There are more complex formulas to consider the retransmission timeout value, and number of packets acknowledged by each ACK and the maximum size of the congestion window.

\subsubsection{Multicast TCP-Friendly}
\begin{figure}
	\centering
	\begin{tikzpicture}
	\path (0,0)  node(ag) []   {}
	(4.3,4) node(a) [rounded rectangle, draw, fill=green!30]   {TCP Friendly CC} 
	(4.3,3.3) node(aux1) []  {} 
		
	(2,2.4) node(b) [rounded rectangle, minimum width=2cm,draw,fill=green!60]  {Single Rate} 
	(2,1.7) node(aux2) []  {} 
	
	(6.5,2.4) node(c) [rounded rectangle, minimum width=2cm,draw,fill=green!60]  {Multi Rate} 
	(6.5,1.7) node(aux3) []  {} 
	
	(1,1) node(d) [rounded rectangle, minimum width=1.3cm,draw,fill=green!60]  {Rate } 
	(3,1) node(e) [rounded rectangle, minimum width=1.3cm,draw,fill=green!60]  {Window } 

	(5.3,1) node(f) [rounded rectangle, minimum width=1.3cm,draw,fill=green!60]  {Rate } 
	(7.5,1) node(g) [rounded rectangle, minimum width=1.3cm,draw,fill=green!60]  {Window }

	(1.5,0)      node(d1) [rectangle, minimum width=1.5cm,draw,fill=pink!60] {RAP}
	(1.5,-.5)    node(d2) [rectangle, minimum width=1.5cm,draw,fill=pink!60] {LDA+} 
	(1.5,-1)     node(d3) [rectangle, minimum width=1.5cm,draw,fill=pink!60] {TFRC}
	(1.5,-1.5)  node(d4) [rectangle, minimum width=1.5cm,draw,fill=pink!60] {TEAR}

	(3.5,0)      node(e1) [rectangle, minimum width=1.8cm,draw,fill=pink!60] {RLA,LPR}
	(3.5,-.5)    node(e2) [rectangle, minimum width=1.8cm,draw,fill=pink!60] {MRCP} 
	(3.5,-1)     node(e3) [rectangle, minimum width=1.8cm,draw,fill=pink!60] {NCA}
	(3.5,-1.5)  node(e4) [rectangle, minimum width=1.8cm,draw,fill=pink!60] {pgmcc}

	(6,0)      node(f1) [rectangle, minimum width=1.8cm,draw,fill=pink!60] {RLC}
	(6,-.5)    node(f2) [rectangle, minimum width=1.8cm,draw,fill=pink!60] {FLID} 
	(6,-1)     node(f3) [rectangle, minimum width=1.8cm,draw,fill=pink!60] {LTS}
	(6,-1.5)  node(f4) [rectangle, minimum width=1.8cm,draw,fill=pink!60] {TFRP}
	(6,-2)  node(f5) [rectangle, minimum width=1.8cm,draw,fill=pink!60] {MLDA}
	
	(8,0)      node(g1) [rectangle, minimum width=1.8cm,draw,fill=pink!60] {Rainbow}
	; 
	
	
	\draw [ line width=1]  (a.south) -| (aux1.north);
	\draw [ line width=1]  (b.south) -| (aux2.north);
	\draw [ line width=1]  (c.south) -| (aux3.north);

	\draw [ ->,line width=1]  (aux1.north) -| (b.north);
	\draw [ ->,line width=1]  (aux1.north) -| (c.north);

	\draw [ ->,line width=1]  (aux2.north) -| (d.north);
	\draw [ ->,line width=1]  (aux2.north) -| (e.north);

	\draw [ ->,line width=1]  (aux3.north) -| (f.north);
	\draw [ ->,line width=1]  (aux3.north) -| (g.north);

%
%
	
	\foreach \value in {1,...,4}
	\draw[->] (d.200) |- (d\value.west);
	
		\foreach \value in {1,...,4}
	\draw[->] (e.200) |- (e\value.west);

		\foreach \value in {1,...,5}
\draw[->] (f.200) |- (f\value.west);

\foreach \value in {1}
\draw[->] (g.205) |- (g\value.west);

	\end{tikzpicture}
	
	\caption{TCP Friendly  \cite{Widmer}.} 
	\label{fig_TFRC}
\end{figure}
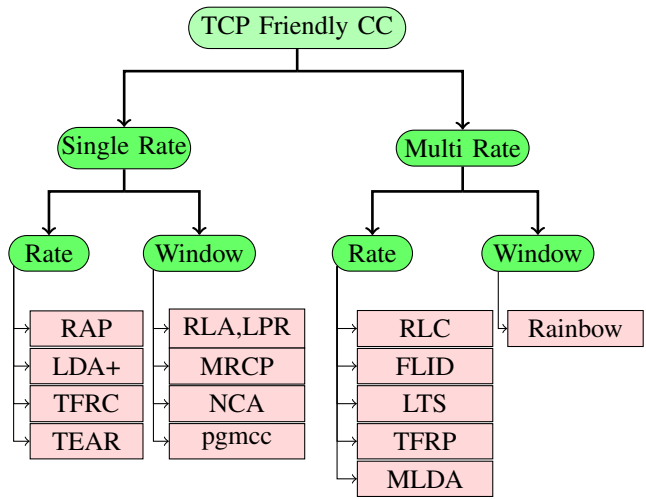

Multicast real time traffic is another main characteristic of the real time flows. For example many users could watch the same Netflix video at the same time. Therefore congestion control for the multicast real time traffic is also required to be considered. Figure \ref{fig_TFRC}. shows friendly classification of single and multi rate.

Rate Adaptation Protocol (RAP) is an end-to-end TCP-friendly transport protocol. It is designed for video-on-demand servers to provide real time applications for the large number of users \cite{Rejaie1998}. RAP is a rate-based congestion control mechanism that is suited for unicast playback of real time streams. RAP adopts an AIMD algorithm for rate adaptation to be fair to use with TCP flows. SCP \cite{Cen1998}  is also proposed for real time streaming of continuous multimedia across the internet. SCP fairly shares the network resources with other TCP and SCP flows. Additionally SCP stream flows smoothly and ensures low and predictable delay.

Like the other transport control protocols,  TCP-friendly algorithms also need to address three issues:  the decision function, the increase/decrease algorithm, and decision frequency. Authors in  \cite{Widmer} surveyed the TCP-Friendly Congestion Control mechanisms.

loss-delay based adaptation algorithm (LDA+) uses the real-time transport protocol (RTP) for collecting loss and delay statistics which are then used for adjusting the transmission behavior of the senders \cite{Sisalemc}.  Multimedia multicast scalability issues of TCP led to TCP Emulation at Receivers (TEAR). TEAR shifts most of the flow control mechanisms to receivers \cite{Rhee2000}. TEAR results in superior fairness to TCP with significantly lower rate fluctuations than TCP. TEAR’s sensitivity to feedback intervals is shown to be very low. It is also shown that even under high feedback latency, TEAR flows exhibit acceptable performance in terms of fairness, TCP-friendliness, and rate fluctuations.

A key issue in the design of source-based multicast congestion control schemes is how to aggregate loss indications from multiple receivers into a single rate control decision at the source. Bhattacharyya et al. in \cite{bhattacharyya2001novel} proposed a novel loss indication filtering called the linear proportional response (LPR) approach.

Authors in  \cite{Rhee} proposed another CC scheme as a large-scale reliable multicast. They incorporates several novel features: (1) hierarchical congestion status reports that distribute the load of processing feedback from all receivers across the multicast group, (2) the relative time delay (RTD) concept which overcomes the difficulty of estimating round-trip times in tree-based multicast environments, (3) window-based control that prevents the sender from transmitting faster than packets leave the bottleneck link an the multicast path through which the sender's traffic flows, (4) a retransmission window that regulates the flow of repair packets to prevent local recovery from causing congestion, and (5) a selective acknowledgment scheme that prevents independent (i.e., non-congestion-related) packet loss from reducing the sender's transmission rate.

Scalable and reliable multicast systems require the consideration of some problems such as  feedback implosion, retransmission scoping, distributed loss recovery, and CC. Therefore authors in \cite{Kasera2000} proposed active services at strategic locations in the networks. They exploit physical hierarchy for feedback aggregation and effective retransmission scoping with minimal router support.

pgmcc is also proposed as a single rate multicast CC scheme \cite{Rizzo2000}. pgmcc achieves scalability, stability and fast response to stochastic network behaviour. The innovation of  pgmcc is a fast and low-overhead procedure to select the acker.

Thus far we studied single rate TCP friendly methods. Following multi rate TCP-friendly  will be discussed in more detail. In \cite{Vicisano} the Receiver-Driven Layered CC (RLC)  is proposed which is receiver driven and requires no per-receiver status at the sender. This helps to overcome scalability problems. Yano et al. in  \cite{Yano2000a}, proposed another approach in which each receiver maintains its own congestion window and individually runs window control modeled after TCP. Other multi rate TCP friendly algorithms have presented  in \cite{Byers2002}, as Fair Layered Increase/Decrease with Dynamic Layering (FLID-DL), \cite{Turletti} as Layered Transmission Scheme LTS, \cite{Sisalemb} as  Multicast Loss-Delay Based Adaptation Algorithm (MLDA), \cite{WaiTian1999} as video multicast using hierarchical forward error correction technique. In \cite{Pan2000}, other TCP friendly algorithms are proposed.

\subsection{Multipath} \label{sec_multipath}
The idea of multipath transport protocols has evolved since the beginning of various Internet connections on mobile devices such as wifi and 4G adapters together. Later on this idea extends to other networks which have multiple connections at the same time such as data center networks. Multipath transport provides attractive features of increased reliability, throughput, fault tolerance, and load balancing capabilities. As a downside multipath stability,convergence, and packet reordering  are the main challenges of multipath transport control. Delay based is applied CC in multipath scenarios \cite{Cao2012}. TCP fairness is another key feature that needs to be taken into account to design multipath TCP\cite{becke2012fairness,Wischik2011a}. Authors in \cite{Chihani-DenisCollange,Habib2016,ref_CC_Multipath_survey} surveyed different multipath TCP protocols.

As a first attempts Equal-Cost Multi-Path routing (ECMP) applied in network layer traffic engineering to achieve load balancing \cite{thaler2000rfc2991}. In 2009 Honda et al. published the Equally-weighted TCP (EWTCP) \cite{honda2009multipath,peng2014multipath}. EWTCP  applies TCP-NewReno algorithm on each route independently of other routes. In receiving Ack, It adjusts the window on multiple routes as $w_r = w_r +a/w_r$ . On each loss, it adjusts the window on multiple routes as $w_r =  w_r/2$.

Decoupled and coupled are two ways to classify multipath congestion control mechanisms. If multipath is a simple extension of a single path TCP (like the previous ones) is called decoupled multipath CC. However they are unfair to single TCP flows. A coupled multipath correlates all the sub-path congestion windows and controls the total congestion window size. Table \ref{tab:multipath}. shows some of the multipath TCP mechanisms

 In 2012, Khalili et al. showed that MPTCP users could be excessively aggressive towards TCP users \cite{Khalili2012}. They design opportunistic linked increases algorithm (OLIA) which is a window-based congestion-control mechanism. OLIA  couples the additive increases and uses unmodified TCP behavior in the case of a loss. Zhang et al. also proposed the weighted ECMP, and they develop a model to obtain the split ratios such that the overall network end-to-end delay is optimized \cite{Zhang2012}. They also applied a flow-based routing model rather than destination-based forwarding to IP networks. Le et al. published the idea of multipath binomial CC \cite{Le2012}. They extend the model of binomial algorithms for single-path to support the concurrent transmission of packets across multiple paths. MultiPath TCP (MPTCP)\cite{ford2013tcp}  adds an extra MPTCP layer which provides CC on the top of multiple TCP subflows. MPTCP starts with one path and increases the number of paths through MP-CAPABLE handshake at initialization and MP-JOIN negotiation.  Web performance could improve if it runs over MPTCP. Authors in \cite{Han2015} studied losses and insufficient band-width utilization of the web over MPTCP.


Web applications are another great usage of the multipath TCP. Object multiplexing design in HTTP/2 improves the web performance. To improve the performance even further \cite{Qian2015} proposed the TM 3 multiplexing architecture. Leveraging multiple concurrent multiplexing pipes in a transparent manner, TM 3 eliminates various types of head-of-line blocking. Energy-Aware Multi-Path TCP is another key point in mobile devices to consider, therefore \cite{Lim2015} proposed eMPTCP which seeks to reduce power consumption compared to standard MPTCP. eMPTCP utilizes a combination of power-aware subflow management and delayed subflow establishment. Wang et al. \cite{Wang2018} published another effort on energy efficient CC for multipath TCP in heterogeneous networks.

Hesmans et al. in \cite{Hesmans2015} proposes a multipath TCP path manager (SMAPP) that delegates the management of paths to the applications. Path quality based multipath AIMD (AIMD-PQ) is proposed in \cite{JiaZhao2015}. AIMD-PQ leverages a combination of packet loss and path quality signals to balance the loads among its sub-paths. The internet policy drop can be a simple way for network operators to reduce the traffic they carry. \cite{popovici_exploiting_2016} studied the policy drop effect on single path and multipath TCP. While monitoring the loss rate and received throughput, end users can detect policy drop detection. Drop based MPTCPs suffer from policy drop unless they design to work separate and independent CC in each path.  In 2013 (combining the functions of HTTP/2, TLS, and TCP directly over UDP) Quick was designed and developed by Google. Quick reduces the client-server session establishment.  \cite{DeConinck2017} proposed the Multipath QUIC (MPQUIC) to use different paths such as WiFi and LTE on smartphones. Complex interdependencies between the multi path losses, packet reordering due to heterogeneous wireless channel features, errors, and link layer retransmissions, as well as their (joint) influence on MPTCP's control mechanism necessitate the comprehensive multipath CC design approach. While relying on a parallel queueing model, Pockrel et al.  \cite{Pokhrel2018} designed a MPTCP algorithm to exploit the route heterogeneity.

Gigabit 5G networks and their high throughput/delay sensitive applications such as remote control of unmanned terrestrial and aerial vehicles implies the need for reliable and latency-constrained data flows. Chiariotti et al. \cite{Chiariotti2018} proposed Latency-controlled End-to-End Aggregation Protocol (LEAP). They forecast the stochastic of the latency-constrained capacity of each path to maximizing the throughput.  Cross-path payload encoding is also applied to control the minimum probability of timely data delivery within a predefined deadline.

\begin{table}[]
	\centering
	\resizebox{0.45\textwidth}{!}{%
		\begin{tabular}{|l|c|c|l|}
			\hline
			Algorithm             & \multicolumn{1}{|p{2cm}|}{\centering Coupling of  \\ subflow \\ windows} & \multicolumn{1}{|p{2cm}|}{\centering TCP \\ Friendly \\ condition}& \multicolumn{1}{|p{2cm}|}{\centering Congestion \\ Indication \\ Signals} \\ \hline
			TCP- new Reno \cite{floyd2004rfc3782}         & No                                               & No                                          & Loss Based                    \\ \hline
			EWTCP \cite{honda2009multipath}                & Yes                                              & Yes                                         & Loss Based                    \\ \hline
			mVegas \cite{vo2014multi}               & Yes                                              & Yes                                         & Delay Based                   \\ \hline
			mReno \cite{vo2014multi}                & Yes                                              & Yes                                         & Loss Based                    \\ \hline
			MSTCP \cite{kokku2007multipath}                & No                                               & No                                          & Loss Based                    \\ \hline
			semi coupled \cite{wischik2011design}         & Yes                                              & Yes                                         & Loss Based                    \\ \hline
			OLIA \cite{khalili2013opportunistic}                 & Yes                                              & Yes                                         & Loss Based                    \\ \hline
			Coupled \cite{han2006multi,kelly2005stability}              & Yes                                              & Yes                                         & Loss Based                    \\ \hline
			ecMTCP \cite{le2011ecmtcp}                & Yes                                              & Yes                                         & Loss Based                    \\ \hline
			Tradeoff \cite{peng2013multipath}           & Yes                                              & Yes                                         & Loss Based                    \\ \hline
			CMP-AIMD \cite{zhou2013study}             & Yes                                              & Yes                                         & Loss Based                    \\ \hline
			Weighted Vegas \cite{cao2012delay}       & Yes                                              & Yes                                         & Delay Based                   \\ \hline
		\end{tabular}%
	}
	\caption{Multipath TCP methods \cite{ref_CC_Multipath_survey}.}
	\label{tab:multipath}
\end{table}
As discussed before, stability of the multipath CC (load balancing) is a real concern to balance the traffic fluctuation between multi paths. Therefore different types of schedulers are proposed to overcome this problem. Network congestion agnostic/aware are the two groups of active traffic scheduling research. Besides, fine/coarse level scheduling is another type of scheduling research classifications. In section\cref{sec_scheduling} scheduling based multipath CC will be discussed more.

\subsubsection{Scheduling Based CC}\label{sec_scheduling}  
Network traffic congestion could be also controlled through multipath scheduling. Therefore different types of scheduling are proposed to alleviate the network load balance problem \cite{Cegarra2012}. Network congestion agnostic or aware are the two classes of schedulers that have been researched. To enable testing of different schedulers for Multipath TCP Paasch et al. \cite{paasch2014experimental}  designed and implemented a generic modular scheduler framework.

\paragraph{Congestion Agnostic/Aware Flow Scheduling} 
Equal Cost Multi Path (ECMP) as one of the congestion agnostic traffic scheduling,  deploys the packet header fields hashing to find the output path. Therefore, collision may happen because of the nature of the hashing functions. Depending on traffic flow size / delay sensitivity distribution,  hash collision could be problematic which will be more discussed in section \cref{sec_DC}.

\paragraph{Homogeneous/Heterogeneous network  flow scheduling}
Data center networks are more homogeneous while the combination of Wifi and cellular networks are heterogeneous multipaths networks. Therefore two different types of schedulers could achieve higher load balancing/throughput.  Choi et al. \cite{Choi2017}\ derived the theoretical limit of the achievable aggregate throughput in heterogeneous wireless multipath environments. They proposed an optimal load balancing (OLB) scheduler that approximately achieves the theoretical throughput limit. Lim et al. \cite{Lim2017} propose and implement an MPTCP path scheduler called ECF (Earliest Completion First), that utilizes all relevant information about a path, not just RTT. It is shown that ECF outperforms the default MPTCP in some scenarios in which heterogeneity paths are used. Dong et al. \cite{Dong2018} showed that a default MPTCP is generally beneficial for throughput-sensitive large flows with large numbers of subflows, while it may be harmful for latency-sensitive small flows running over heterogeneous paths. They propose to dynamically adjust the subflows according to application workloads ( DMPTCP).

\paragraph{Fine/Coarse grain scheduling}
Fine coarse grain scheduling has different advantages and disadvantages. Packet , subflow (fixed/variable size), and flow scheduling are different types of schedulers. Packet and subflow scheduling achieve a more balanced load, but they may suffer from packet reordering. Kandula et al. \cite{Kandula2007} show that a single flow can be systematically split across multiple paths without causing packet reordering. They proposed scheduling called FLARE. It attains accuracy and responsiveness comparable to packet switching without reordering packets. A power aware scheduling mechanism in \cite{Lim2015} has also been designed to schedule the subflows. Coflow is another technique that is proposed as mixing more short flows. Some applications deploy collections of parallel short flows (to mix as coflow) to convey job-specific communication .  Optimizing a coflow’s completion time (CCT) decreases the completion time of corresponding job \cite{chowdhury2012coflow,chowdhury2014efficient}. Authors in \cite{Guo2017,Verma2017,8486340,Sinha,Vanini,perry2017flowtune} proposed other granularity scheduling.

\paragraph{Other  Scheduling}
Reactive/Proactive and priority based scheduling are other classes of research. Oh et al. \cite{Oh2015} proposed a scheduling algorithm that performs packet scheduling according to the receiver buffer and network delay.  This method estimates out-of-order packets and assigns data packets to sub-flows based on the estimation. Hong et al.  \cite{hong2012finishing} proposed Preemptive Distributed Quick (PDQ) flow scheduling that is a protocol designed to complete flows quickly and meet flow deadlines. PDQ is implemented in a distributed way to improve the scalability of the protocol.  

Zhang et al.  \cite{zhang16} proposed a Priority-based flow scheduling for online social network datacenters(PFO). Their method allocates rates for different flows based on flow size and deadline information. it decreases the average completion time for bursty flows and ensures a high throughput.

Network Coding (NC) is also a subset of network information theory that has led to advancements in the network throughput optimization.  NC involves performing operations other than mere forwarding and replication at the network nodes. Leveraging a given while encoding and merging the relevant messages at the relay node is one of the operations that NC is used to increase network efficiency. After considering the merging mechanisms, multipath traffic engineering and traffic multicast (or broadcast ) are the two most applicable cases for NCs

Linear NC is the most cited NC approache  \cite{li2003linear}. Linear NC uses a block of data as a vector and a node applies a linear transformation to that vector before passing it on. 



As a tentative future trend, Software Defined Networks  (SDNs) \cite{casado2007ethane, kreutz2015software, nunes2014survey,patel2017survey} are another new research trend to make the networks more programmable. Control and forwarding planes are  separated in SDNs to improve the management. Sonkoly et al. \cite{sonkoly2014sdn} exploit large-scale SDN based testbeds that support multipath transport control protocols. Programming Protocol-independent Packet Processors (P4) programmable data plane is proposed to improve the SDN networks \cite{bosshart2014p4}.  


\subsection{Latency and Throughput  Optimization (L\&T) in Data Center (DC) Networks}\label{sec_DC}
As it is discussed before, different network and traffic characteristics implies customized transport protocol design consideration. A Data Center (DC) is a computing environment with large amounts of inter-connected resources such as computational and storage resources. A typical DC may have hundreds of thousands of servers. Data center resources interconnect using various data communication network topologies. DC network topologies (section \cref{sec:net_topo}) are designed to be resilient\cite{Gill2011,Zhang2017,zhuo_understanding_2017}, massively scalable  \cite{Greenberg2011}\ and multi path\cite{Raiciu2010,zhou2014wcmp}. A typical datacenter is made up of thousands of servers connected with a large network located close to each other in a large hall. Millions of customers' virtual machines should be served in DC networks \cite{He2016}. Mysore et al. \cite{NiranjanMysore2009} proposed Portland as a scalable fault-tolerant layer 2 DC network fabric. PortLand employs a logically centralized fabric manager that maintains a soft state about network configuration information. They deploy positional pseudo MAC addresses for efficient forwarding, routing, and VM migration. Optical interconnects have also gained attention recently as a promising solution offering high throughput, low latency and reduced energy consumption compared to electronic packet networks based on commodity switches. Kachris et al. \cite{Kachris2012} surveyed optical interconnects for DC. Further, traffic flow monitoring in these scalable networks is a challenge to overcome \cite{Li2016}.

Further, a DC traffic pattern needs to meet specified standards (section \cref{sec:traffic}). Traffic that is running in DC networks is a mix of user-generated interactive traffic, traffic with deadlines \cite{vamanan2012deadline,D3_wilson2011better}, and long-running traffic. The most common DC traffic challenges include but are not limited to unpredictable traffic matrix \cite{judd2015attaining}, mix of flow types and sizes, and delay sensitivity. Traffic burstiness, packet reordering, performance isolation, incast problems, and  outcast problems  are some of the most common challenges in designing DC networks and transport protocols. Low latency service is the main requirement of the DC characteristics. To reduce the service delay \cite{liu2013low} surveyed the attempts into four groups of works including reducing queue length, accelerating retransmissions, prioritizing mice flows, and exploiting multi-path \cite{al2010hedera}.

DC network traffic control can be also managed in distributed or centralized fashion. However Distributed methods are more scalable and therefore more interested in practice. Further different types of networks and transport protocols are required in different aspects of the DC networks. Inter-DC  \cite{tao2020congestion} and intra-DC are the two main aspects of DC networks that have their specific characteristics and transport requirements. Jain et al. in  \cite{jain2013b4} utilized SDN networks to design, implement, and evaluate the B4 as a private WAN connecting Google's data centers across the planet. Authors in \cite{Huang2018,Wang2018a,Sreekumari2016,JiaoZhang2013,khan2015handbook,Noormohammadpour2018,xia2016survey}, surveyed DC network topologies, traffic characteristics and transport protocols. \\

\subsubsection{Related works}
\paragraph{Network Topologies} \label{sec:net_topo}
DC network characteristics are different from other networks. DC servers are typically located in rows of racks of servers in a big hall near to each other. There are high speed links and very short distance between the servers. Therefore data is transferred from one server to another in only a few microseconds \cite{SDNChapter}. Any delay such as queuing in this short distance  could be a significant service effect. Besides high speed optical links and slow increase of buffer size are the reason to consider shallow buffer switches in DC network analysis \cite{bai_congestion_2017}. Moreover scalability is a key requirement of the data center networks \cite{vahdat2010scale}. Researchers are studying two types of network topologies as switch centric and server centric types of networks. Zhang et al. \cite{zhang2012architecture}  comprehensive surveyed  research activities in DCNs. They emphasis on the network architecture design, congestion notification algorithms, TCP Incast, and power consumption.

\begin{figure}
		\centering
		\resizebox{.5\textwidth}{!}{%

\begin{tikzpicture}[thick, node distance=1em]

\begin{scope}[every node/.style={draw,rectangle,thick},inner sep=1.0em]
\tikzstyle{level 1}=[sibling distance=6cm,every child/.style  ={line width=4pt},inner sep=0.8em]
\tikzstyle{level 2}=[sibling distance=3cm,every child/.style  ={line width=2pt},inner sep=0.5em]
\tikzstyle{level 3}=[sibling distance=1.5cm,every child/.style={line width=1pt},inner sep=0.1em]

\node [fill=green!60]{Switch}
child {node [fill=green!60]{Switch}
	child {node [fill=green!60]{Switch}
		child {node [draw,rectangle,rotate=90,minimum width=2cm,minimum height=1.5cm,draw,fill=pink!60,inner sep=0.1em] {Rack of Servers}}
	}
	child {node [fill=green!60]{Switch}
		child {node [draw,rectangle,rotate=90,minimum width=2cm,minimum height=1.5cm,draw,fill=pink!60,inner sep=0.1em] {Rack of Servers}}
	}
}
child {node[fill=green!60] {Switch}
	child {node [fill=green!60]{Switch}
		child {node [draw,rectangle,rotate=90,minimum width=2cm,minimum height=1.5cm,draw,fill=pink!60,inner sep=0.1em] {Rack of Servers}}
	}
	child {node  [fill=green!60] {Switch}
		child {node [draw,rectangle,rotate=90,minimum width=2cm,minimum height=1.5cm,draw,fill=pink!60,inner sep=0.1em] {Rack of Servers}}
	}
}
;
\end{scope}

\end{tikzpicture}
}
\caption{Fat Tree topology, higher bandwidth links to the core switch}
\label{fig:fattree}
\end{figure}
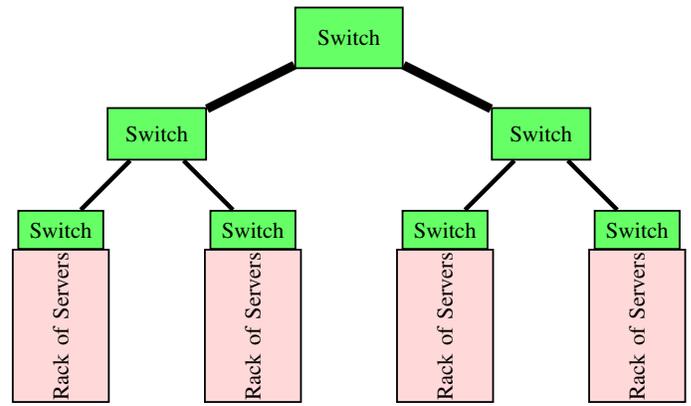

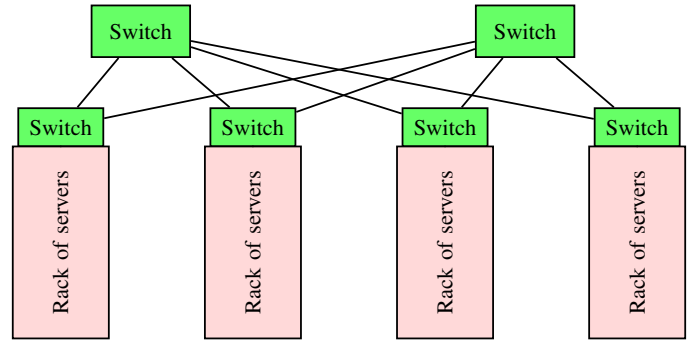
\begin{figure}
			\centering
	\resizebox{.5\textwidth}{!}{%

\begin{tikzpicture}[thick, node distance=1em]

\foreach \n in {0,...,1}{
	\node (level 1 \n) at (2.0cm + \n*4*1.5cm,3.0) [draw,rectangle,inner sep=0.8em,fill=green!60] {Switch};
}

\foreach \n in {0,...,3}{
	\node (level 2 \n) at (0.75cm + \n*2*1.5cm,1.5) [draw,rectangle,inner sep=0.5em,fill=green!60] {Switch};
	
	\draw [thick] (level 2 \n) to (level 1 0);
	\draw [thick] (level 2 \n) to (level 1 1);
}

\foreach \n in {0,...,3}{
	\node (node \n) at (0.75cm + \n*2*1.5cm,-.3) [draw,rectangle,rotate=90,minimum width=3cm,minimum height=1.5cm,draw,fill=pink!60,inner sep=0.1em] {Rack of servers};
	
	\draw [thin] (node \n) to (level 2 \n);
}

\end{tikzpicture}
}
\caption{Leaf Spine}
\label{fig:leafspine}
\end{figure}

\paragraph{Switch Centric}
These types of networks are  multi-rooted tree topologies, interconnecting the edge, aggregate, and core switches. Fat-Tree (Figure \ref{fig:fattree}) is one of the most popular switch centric networks \cite{Fares2008}. Leaf-Spine (Figure \ref{fig:leafspine})  is another very common topology \cite{Alizadeh2013LeafSpine,alizadeh2013data}.

\paragraph{Server Centric}
In Server centric networks, servers with multiple network ports act as not only end hosts, but also relay nodes for other servers. These networks are more scalable and cheaper than switch centric. However, in these networks, servers have to spend some portion of their processing power on rerouting the incoming traffics.

BCube is one of the server centric network topologies that is specifically designed for shipping-container based, modular data centers\cite{Guo2009} . Dcell is another server centric network that scales doubly exponentially as the node degree increases \cite{Guo2008}. Jellyfish is more cost-efficient than a fat-tree, supporting as many as 25\% more servers \cite{Singla2011}. Jellyfish also allows great flexibility in building networks with different degrees of oversubscription. BCube is another server centric data center topology \cite{Huang2011}\\

\paragraph{DC Traffic Patterns and problems}\label{sec:traffic}
Traffic patterns in DC networks consist of mice and elephant flows. Mice and elephant  flow proportion varies based on the type of applications that the DC is serving \cite{Noormohammadpour2018}.  Besides mice flows are typically delay sensitive while elephant flows are throughput sensitive traffic. Alizadeh et al.  \cite{Alizadeh2012Less_is_More} proposed an idea to trade a little bandwidth for ultra-low latency in the DC. East-west vs south-north traffic is another classification in DC network traffic. To serve delay sensitivity, Preemptive Distributed Quick (PDQ) \cite{hong2012finishing} is one of the techniques that schedule the flows to meet the flow deadlines PDQ. Chen et al.  \cite{chen2016scheduling} proposed a scheduling algorithm that deadline flows minimally impacting the FCT of non-deadline flows. It is more challenging to have reactive CC mechanisms meet the flow deadline because of stringent latency requirements to cloud data center networks. Therefore proactive methods are also increasing exponentially \cite{hu_augmenting_2018}. Authors in \cite{ChengRJQZS16,roy2015inside,network_traffic_chars} studied traffic characteristics of the different DC networks. Besides, one to many, many to one are other types of traffic characteristics that are mostly common in DC networks.

\paragraph{Performance Isolation}
DCs are the main infrastructure to provide cloud services. Sharing network resources among cloud tenants is the technique to increase the resource utilization, customers, and revenue consequently. Performance isolation prevents selfish tenants from controlling and not overusing the resources \cite{ChengRJQZS16}. EyeQ \cite{Jeyakumar2013}  provides tenants with bandwidth guarantees to isolate the network performance at the edge.

\begin{figure}
 	\centering
 	\includegraphics[width=0.5\textwidth]{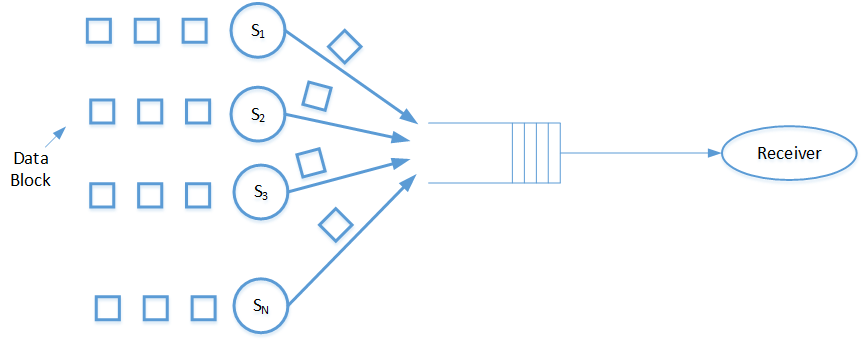}
 	\caption{Incast problem in Data Center networks.}
 	
 	\label{fig:Incast}
 \end{figure}

\paragraph{Incast problem}
 In DC networks Incast is the consequence of the various synchronized many-to-one communication patterns created by partition/aggregate applications. Some of these applications are web search, MapReduce, social network content composition, and advertisement selection. Incast causes the queue buildup and buffer pressure in switches \cite{Thiruvenkatam, Vasudevan2009,Tahiliani2015}. According to the reports from large data center operators such as Facebook \cite{roy2015inside}, Google \cite{singh2015jupiter}, and Microsoft \cite{kandula2009nature},  incast problems mostly occur close to the receiver (Top of Rack switches). Figure \ref{fig:Incast}. shows the  Incast problem.

Researchers have published various attempts to overcome the incast problem. Alipio et al.  \cite{alipio2019tcp} reviewed the incast problem  solutions that are provided. They classified the solutions as TCP based and non-TCP based techniques. They classified non-TCP approaches further as link \cite{Alizadeh2008,devkota2010performance, Kabbania2010, Ruan2017}, application \cite{podlesny2011application}  and cross layer \cite{Tseng2017} techniques. Additionally the TCP based approaches are also classified into active queue management \cite{D3_wilson2011better,hong2012finishing}, delay based, and multipath TCP (section \cref{sec:DC:multipath}) group of works.

Alizadeh et al.  \cite{Alizadeh2008,ruan2020nonlinear} described and analysed the Quantized Congestion Notification (QCN) as an attempt to develop an ethernet congestion control algorithm for hardware implementation. Random TCP timeout is also one of the methods to reduce the buffer pressure and incast problems consequently. Fair Quantized Congestion Notification (FQCN) \cite{Zhang2011} is proposed to improve fairness of multiple flows sharing one bottleneck link. Incast-Avoidance TCP (IA-TCP) \cite{Hwang2012} is also proposed to avoid the TCP incast congestion problem.  IA-TCP employs the rate-based algorithm at the aggregator node, which controls both the window size of workers and ACK delay. Deadline-Aware Datacenter TCP (D2TCP) \cite{Vamanan2012}, as a distributed and reactive approach, employs another congestion avoidance algorithm. D2TCP utilizes the ECN feedback and deadlines to modulate the congestion window via a gamma-correction function. It is also shown that \cite{Wu2012} a proper tuning of ECN at the intermediate switches can alleviate Incast problems. Authors in \cite{JunZhang2013,Das2013,JiaoZhang2013a,Cheng2013,huang2015packet,Hwang2014} proposed other approaches to ameliorate the incast problem.

Vasudevan et al. \cite{Vasudevan2009} proposed a safe and effective fine-grained TCP retransmissions for DC networks. RCC \cite{Xu2017} is a receiver side Incast solution that makes effective use of a centralized scheduler and Explicit Congestion Notification (ECN) at the receiver. Wu et al.  \cite{Wu2010} designed an Incast congestion Control for the TCP (ICTCP) scheme at the receiver side to adjust TCP receive window proactively.  Wang et al. \cite{wang18} also proposed proactive Incast congestion control in a DC network serving web applications. Packet pacing is also proposed as another Incast solution.  Rezaei et al. \cite{Rezaei2019} proposed ICON as an Incast solution using fine grained control over sending rate by pacing traffic in data center Networks.

Incast scalability is another issue to overcome recently. NDP re-architects the datacenter networks and stacks for low latency and high performance. This method uses back pressure along with trimming the packet and forwarding the header fields to overcome reordering problems in the receiver\cite{Handley2017}. NDP can deal  with massive incast events and will prioritize traffic from different senders on RTT timescales. In NDP Incast performance is compared with other methods when the number of senders increases up to 432-nodes in FatTree topology. In spite of all great advantages of NDP, it is not backward compatible with commodity switches. Therefore it is an expensive and time consuming solution to develop.

\paragraph{Outcast problem}
Prakash et al. \cite{prakash2012tcp} presents that outcast problems occur when two conditions are met. First switch uses taildrop queue discipline, and second a large set of flows and a small set of flows arriving at two different input ports compete for a bottleneck output port at a switch. it happens  in DCs because of their many to one traffic pattern.


\paragraph{Related studies}
Network programmable data planes increase flexibility and efficiency of the network. In-Band Network Telemetry is a recent advanced feature of the programmable networks \cite{Mari2018}. Probabilistic In-band Network Telemetry (PINT) \cite{basat2020pint} is an in-band network telemetry with less overhead cost. HPCC \cite{li2019hpcc} leverages in-network telemetry (INT) to obtain precise link load information and controls traffic precisely.

Data Plane Development Kit (DPDK) \cite{dpdkonline} is an open source software project that is managed by the Linux Foundation. DPDK provides a set of data plane libraries and network interface controller polling-mode drivers. It offloads TCP packet processing from the operating system kernel to processes running in user space. This improves the  computing efficiency and higher packet throughput than is possible using the interrupt-driven processing provided in the kernel. This technology is mostly applied in software switches. To reduce the delay of short flow processing,  mTCP \cite{jeong2014mtcp}, a high-performance TCP stack is designed for multicore systems implemented in user- space. mTCP is designed to 1- translate multiple expensive system calls into a single shared memory reference, 2- allow efficient flow- level event aggregation, and 3- perform batched packet I/O for high I/O efficiency.

\subsubsection{Data Center  Approaches}
DC traffics are very high speed and are volatile; therefore centralized schemes are not easy to scale and typically suffer from delays and distributed mechanisms seem more desirable .

\paragraph{Centralized}
Centralize CC is one group of studies to control the network traffics and congestion. A central scheduler in Hedera \cite{al2010hedera} aims to assign flows to non-conflicting paths. Flowtune \cite{perry2017flowtune} is another centralized approach that utilizes an allocator to control the flowlets in DC Networks. \\

\paragraph{Distributed}
Distributed control is the most popular control system in any scalable systems. Thus far many distributed mechanisms have been proposed in DC networks. Some of these mechanisms are relatively cheap to develop through modifying the end host  \cite{Alizadeh2010,jiang_network_2015}. However some others need hardware change on middle switches, which makes them more expensive \cite{Handley2017,Alizadeh2013}. Hardware processing can also reduce the processing time in the end host NIC \cite{arashloo_hotcocoa:_2017}. Data Center TCP (DCTCP) \cite{Alizadeh2010,Alizadeh2011}, Leveraging ECN, DCTCP  is able to gauge the extent of congestion and therefore DCTCP achieves adaptive rate reduction rather than halving the transmission rate in response to network congestion.

As it is discussed in transport protocols of the previous networks, DCs can also rely on delay or ECN to control their network congestion \cite{Zhub2016}. However the problem with delay based is that a very fine grain clock is required in the end host to keep track of the packet delays. High speed and low distance traffic paths take about microsecond delay to transfer the traffics. Therefore it is more challenging to achieve high performance in DC networks with delay based transport protocols. Authors in \cite{Wang2015,mittal2015timely,Ghorbani2017,Zhang2017a,Cho2017,Montazeri2018} proposed different delay based flow control in DC networks.

ECN based transport protocols are another group of DC transport protocols. ECN is utilized in different ways such as dynamic ECN marking threshold \cite{Lu2018}, enabling ECN in multi-service multi-queue \cite{bai2016enabling}, adaptive marking threshold method for delay-sensitive \cite{Zhang2016} traffics, Time-based Congestion Notification (TCN) \cite{Bai2016}

Back pressure is another technique that is also tried in DC networks. Priority-based Flow Control (PFC) uses a back pressure mechanism to provide loss less transmission. However PFC is a coarse grain mechanism because it works based on ports and is not flow based. Therefore it consequently causes congestion spreading and poor performance \cite{Zhua2015}. In PFC the congested switch sends PAUSE/RESUME messages to uplink entities to stop/resume sending traffic on that link. A drawback of PFC is that it is congestion spreading. 

As it is discussed in DC network topologies and their traffic characteristics, minimizing the switch buffers as a shallow buffer boosts the performance of DC networks. Therefore pFabric is proposed as a near-optimal datacenter transport \cite{Alizadeh2012,Alizadeh2013}.  pFabric provides near theoretically optimal flow completion times even at the 99th percentile for short flows, while still minimizing average flow completion time for long flows. As  pFabric requires specialized hardware, thus pHost \cite{Gao2015} proposed to decouple the network fabric from scheduling decisions. pHost introduces a new distributed protocol that allows end-hosts to directly make scheduling decisions.

While leveraging underlying complementary strategies of pFabric and PDQ \cite{hong2012finishing}, PASE is proposed as a transport framework that synthesizes the PDQ and pFabric. PASE adjust the endpoints in network prioritization (used in pFabric), and arbitration (used in PDQ).\\

\paragraph{Multipath} \label{sec:DC:multipath}
Considering the most popular DC network topologies as switch centric networks, multipath transport protocol is a common approach. Thus far different granularity scheduling have been applied in multipath DC research. As different granularity multipath approaches packet based load balancing \cite{Cao2013lb}, flow level adaptive routing \cite{Kabbani2014, yang2020qcluster},  fixed size flowlet called flow cell \cite{he2015presto},and variable size flowlet granularity \cite{AlizadehFlowlet2016} have been published.  Random Packet Spraying (RPS) is another approach that is proposed in \cite{dixit2013impact}. They argue that symmetrical DC network topologies provide the multiple equal-cost paths between two hosts. These paths are composed of links that exhibit similar queuing properties. Therefore, TCP is able to tolerate the induced packet reordering and maintain a single estimate of RTT.

Cao et al. \cite{Cao2013} proposed  Explicit Multipath (XMP)  Congestion Control  through Buffer Occupancy Suppression (BOS) algorithm, and Traffic Shifting (TraSH). BOS helps as controllable link buffer occupancy to meet the low latency requirement of small flows. TRaSh shifts the multipath traffic, thus improving the throughput of large flows.

TinyFlow \cite{xu2014tinyflow} breaking elephants flows down into mice and applies ECMP. CONGA \cite{Alizadeh2014} is also a distributed congestion-aware load balancing mechanism. CONGA splits the TCP flows into flowlets, estimates real-time congestion on fabric paths, and allocates flowlets to paths based on feedback from remote switches. DC networks serving cloud tenants could apply transport protocols on server hypervisors. Authors in \cite{Cronkite-Ratcliff2016, he2015presto} present virtualized congestion control  and Presto as edge-based load balancing. Chen et al. \cite{chen2016fast} present that loss is not uncommon in DC networks. They propose a loss recovery approach FUSO that exploits multi-path diversity in DCN for transport loss recovery. EMAN \cite{zhangy2020achieving} distributes outbound traffic from a sending end host depending on the bandwidth of each path. EMAN uses feedback from the receiver to update each path’s available bandwidth. Then it reacts to network asymmetries caused by failure and flow competition.

 PASE \cite{munir2015friends} is also designed as a transport framework that synthesizes  transport strategies such as self-adjusting endpoints (used in TCP style protocols), in network prioritization (used in pFabric), and arbitration (used in PDQ). PASE is backward compatible, and it does not require any changes to the network fabric.  PIAS \cite{bai2015information} is a DCN flow scheduling mechanism that tries to minimize FCT by mimicking Shortest Job First (SJF). PIAS is one of the flow size-agnostic scheduling for commodity data centers. PIAS emulate SJF without knowing flow sizes in advance. PIAS implements a Multiple Level Feedback Queue (MLFQ) based on multiple priority queues that is available in existing commodity switches. PIAS uses the MLFQ to gradually downgrade a flow from higher-priority queues to lower-priority queues. Downgrading method is based on the number of bytes it has sent.  Therefore short flows are likely to be finished in the first few high-priority queues and thus be prioritized over long flows in general.

\paragraph{TCP Adaptive Initial Rate }
TCP initial adaptive rate improves the performance of applications in which there are a myriad of short flows. Short flows may finish before reaching the maximum rate and exploring the available resources. Padmanabhan et al. proposed \cite{padmanabhan1998tcp} that the sender caches network parameters to avoid paying the slow start penalty for each web page download. Paced Start \cite{hu2003improving} uses the difference between data packet spacing and the acknowledgement spacing to improve TCP startup. Hoe et al.  \cite{hoe1996improving} shows another initial rate improvement.

In \cite{ruth2018demystifying, ruth2019tcp} authors study the initial window configurations of major Content Delivery Networks(CDN). They also found a high amount of initial window customization that is beyond current Internet standards. Studies in \cite{dukkipati2010argument} show that increasing the window size from four to ten segments improves the HTTP response up to ten percent.  Radhakrishnan et al. \cite{radhakrishnan2011tcp} published the TCP fast open mechanism that data exchange during TCP’s initial handshake. They applied a token that verifies IP ownership to address the security issue. Kodama et al. \cite{kodama2011initial} proposed the Initial Window adaptation through Rate-Based Pacing (RBP) start method.  RBP algorithm is applied here for only one RTT after the connection establishment.

Considering the sequence of start up rate adaptation, reinforcement learning is another technique to apply in startup rate adaptation. In \cite{nie2019dynamic} group-based reinforcement learning (RL) is developed to enable a web server, through trial-and-error, and to adaptively set a proper IW for a web flow. Initial Congestion Window adaptation is also tried for 5G Mobile Edge Computing through Deep Reinforcement Learning \cite{xie2019adaptive}.

Despite the previous studies to find the maximum possible startup rate, there are studies tried to reduce the startup rate. Gentle slow start \cite{memon2019gentle} is a method that tries to alleviate the incast problem in DC networks.

\subsubsection{Future Trends}

  \paragraph{Advanced Network Interface Cards (NICs)}
Offloading the packet processing tasks from general-purpose CPUs is one of the main purposes to develop the advanced Network Interface Cards. These cards are designed to support both network virtualization and application-specific tasks.  Grant et al. \cite{grantsmartnic}  introduced FairNIC, a system to provide performance isolation between tenants utilizing the full capabilities of a commodity system-on-a-chip (SoC) SmartNIC. Modern smart NICs have fully programmable and energy-efficient multi-core processors \cite{firestone2016smartnic,le2017uno} to increase their flexibility and efficiency.

\begin{figure}
	\includegraphics[width=.5\textwidth]{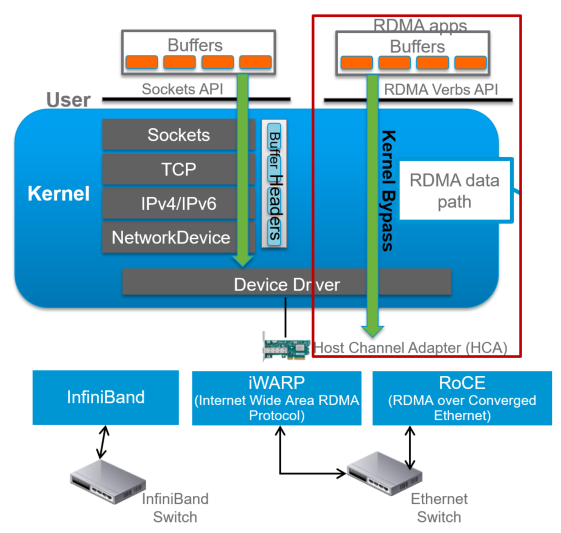}
	\caption{RDMA \cite{RDMAonline}}
	\label{fig:RDMADiagram}
\end{figure}

\paragraph{RDMA, Kernel Bypass} \label{sec:RDMA}
 Over the recent years 40/100/400 Gbps links have been growing rapidly \cite{yuanwang2020100g, maniloff2019400g}.  Besides the classic disk I/O bottleneck is migrating to solid-state disk or in-memory computing \cite{ousterhout2010case}. Therefore CPU consumption and packet processing latency of traditional TCP/IP stack become significant.  As it was discussed before, delay sensitive applications can't tolerate this much delay on DC massive networks. Remote Direct Memory Access (RDMA) is developed as a direct memory access from the memory of one computer into that of another without involving either one's operating system (Figure \ref{fig:RDMADiagram}). This enhances  high-throughput and low-latency networking, which is especially beneficial in enormous parallel computer clusters. RDMA may soon replace TCP in data centers  \cite{mitchell2013using,dragojevic2014farm,Kalia2014}. Initial RDMA could be applied on lossless links like infiniband networks.

InfiniBand (IB) is a communications standard used in high-performance computing that features very high throughput and very low latency. InfiniBand has been applied in either a direct or switched interconnect between servers and storage systems, as well as an interconnect between storage systems. InfiniBand provides RDMA capabilities for low CPU overhead. As an interconnect IB competes with Ethernet. Gusat et al. \cite{gusat_congestion_2005} present the infiniband CC mechanism that consists of congestion detection/marking, congestion signaling, injection rate reduction, and injection rate recovery. Hutchison et al. \cite{hutchison_high_2005,Yu2008} presents RDMA performance on InfiniBand clusters and WAN. Jiang et al. \cite{Jiang2018} surveyed the link layer congestion management of lossless switching fabric.

RDMA is applied to improve the performance of various distributed applications. Ouyang et al. \cite{ouyang_high_2011} proposed the Pipelined Process Migration with RDMA(PPMR) to overcome inefficient I/O overhead. They achieve a 10.7X speedup to complete a process migration over the conventional approach. Ren et al. \cite{Ren2012} designed a middleware to transfer data over the RDMA. Their design includes resource abstraction, task synchronization, and scheduling. Their results outperforms the classical approach while maintaining very low CPU consumption. Kalia et al. \cite{Kalia2014} designed and implemented the HERD which is a key-value system over the RDMA networks. Their design uses a single round trip for all requests and supports up to 26 million key-value operations per second with $5\mu s$  average latency. Binnig et al. \cite{binnig2015end} design a distributed in-memory Database Management Systems (DBMS) over the RDMA networks. Paxos helps State Machine Replication (SMR) to tolerate various types of failures.  APUS \cite{wang_apus:_2017} is an RDMA-based Paxos protocol that aims to be fast and scalable.
Chiller: Contention-centric Transaction Execution and Data Partitioning for Fast Networks \cite{zamanian_chiller:_2018}.

Latency reduction has also been tried in different aspects of the networks. Sriraman et al. \cite{sriraman_deconstructing_2017} observe extreme tails latency that happens in UDP/TCP-IP,  but not in RDMA packet exchanges. A zero copy transport \cite{yi_towards_2017} is another approach that is designed for distributing dataflow frameworks. Zero copy unifies application and networking buffer management and completely eliminates unnecessary memory copies.

RDMA congestion control is also a growing research challenge over the recent years. RDMA provides hop flow control and rate based end to end congestion control \cite{gran2010first}.  Guo et al. \cite{Guo2016} describes the challenges of applying RDMA over commodity Ethernet (RoCEv2). They designed a priority flow-control (PFC) mechanism to ensure large-scale deployment.  They addressed the safety challenges brought by PFC-induced deadlock, RDMA transport livelock, and the NIC PFC pause frame storm problem. Tagger \cite{Hu2017} is also proposed to prevent the deadlock problem in RDMA. Tagger uses a tagging scheme that can be developed to ensure that no deadlock will occur under any failure conditions. 

FaSST is also a RDMA-based system \cite{kalia2016fasst} that provides distributed in-memory transactions with serializability and durability. FaSST has scalable and simple distributed transactions with two-sided (RDMA) datagram RPCs. RoCEv2 relies on Priority-based Flow Control (PFC) to enable a drop-free network. However, head of Line blocking and unfairness are the challenges of PFC. DCQCN is introduced \cite{Zhua2015} as an end-to-end congestion control scheme for RoCEv2. Escudero et al. \cite{Escudero-Sahuquillo2018} proposed Flow2SL-ITh is a congestion control mechanism for infiniband networks. Flow2SL-ITh is a technique that combines a static queuing scheme (SQS) with the closed-loop congestion control mechanism included in IBA-based hardware (a.k.a. injection throttling, ITh). Flow2SL-ITh separates traffic flows in different virtual lanes (VLs) in order to reduce HoL blocking, while the injection rate of congested flows is throttled. ACCurate \cite{Giannopoulos2018} is a congestion control protocol for RDMA Transfers that assigns max-min fair rates to flows. ACCurate is a scheme suitable for efficient hardware implementation

RoCE provides low latency and low CPU usage but it suffers from network stability. The reason is that RoCE utilizes the PFC which is a backpressure mechanism to provide lossless networking to RDMA. 

RoGUE \cite{Le2018} is a  congestion control and recovery mechanism for RDMA over Ethernet that does not rely on PFC. This allows the use of RDMA for high performance, supporting both the Reliable Connection (RC), and Unreliable Connection (UC) RDMA transports.

Incast resolution and multipath transport are two interesting recent RDMA based DC network research. Dart \cite{xue_dart:_2018} is proposed as a divide-and-specialize approach, which separates the  receiver congestion and remaining in-network congestion. For receiver congestion Dart uses direct apportioning of sending rates (DASR) and directs each n senders (that it is receiving from) to cut their rate by a factor of n. For the spatially-localized case Dart provides fast response by adding novel switch hardware for in-order flow deflection (IOFD) because RDMA disallows packet reordering.  Multipath RDMA (MPRDMA) efficiently utilizes the rich network paths in data centers. MP-RDMA proposed in \cite{lu2018multi} employs three  techniques to address the challenge of limited RDMA NICs on-chip memory size: 1) a multi-path ACK-clocking mechanism to distribute traffic in a congestion-aware manner without incurring per-path states, 2) an out-of order aware path selection mechanism to control the level of out-of-order delivered packets, and 3) a synchronise mechanism to ensure in-order memory update whenever needed.

Mushtaq et al. \cite{mushtaq2019datacenter} studied the bulk of research on DC congestion control and found that the switch scheduling algorithm as the most essential feature. They present that if the rate-setting algorithm at the host is reasonable then congestion control mechanisms that use Shortest-Remaining-Processing-Time (SRPT) achieve superior performance. They reach the point that SRPT’s performance is more robust to host behaviors. They observe that approximate and deployable SRPT (ADS) designs exist. ADS leverage the small number of priority queues supported in almost all commodity switches. Their result is close to SRPT and  better than FIFO.

\section{Latency and Throughput Optimization (L\&T)  in Wireless Networks} \label{sec_wireless}
\begin{figure}
	\centering
		\resizebox{.5\textwidth}{!}{%
	\begin{tikzpicture}
	
	\path (0,0)  node(ag) []    {}
	(3.2,3) node(a) [rectangle,draw,fill=green!60]               {Wireless networks, \ref{sec_wireless}} 
	(3.2,2) node(aux1) []               {} 
	
	(0,1) node(b) [rectangle,text width=2.5cm,draw,fill=green!30]               {Cellular, 3G/4G/5G,  \ref{sec_cellular}} 
	(3.2,1) node(c) [rectangle,text width=2.5cm,draw,fill=green!30]               {Fixed adhoc networks, \ref{sec_sensor_Network}} 
	(6.3,1) node(d) [rectangle,text width=2.5cm,draw,fill=green!30]               {Mobile Adhoc Network, \ref{sec_Manet_Vanet}} 
	(6.3,0) node(aux2) []               {} 
	
	(3.2,-1) node(e) [rectangle,text width=2.6cm,draw,fill=green!30]               {VANET, \ref{sec_vanet} } 
	(6.5,-1)  node(f) [rectangle,text width=2.6cm,draw,fill=green!30] 				 {FANET, \ref{sec_fanet}} ;

	\draw [ line width=1]  (a.south) -| (aux1.north);
	\draw [ line width=1]  (d.south) -| (aux2.north);
	
	\draw [->, line width=1]  (aux1.north) -| (b.north);
	\draw [->, line width=1]  (aux1.north) -| (c.north);
	\draw [->, line width=1]  (aux1.north) -| (d.north);
	
	\draw [->, line width=1]  (aux2.north) -| (e.north);
	\draw [->, line width=1]  (aux2.north) -| (f.north);	
	\end{tikzpicture}
}
	\caption{Wireless Section.} 
	\label{fig:secwireless}
\end{figure}
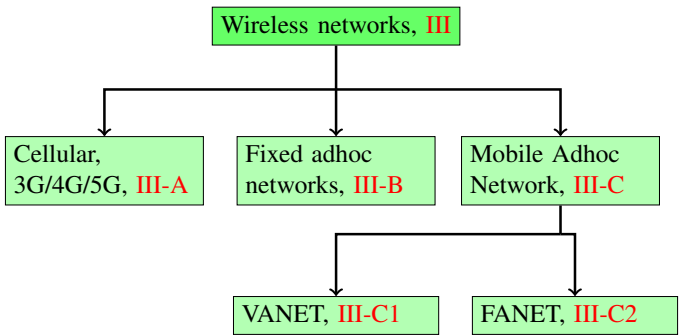

Data loss in wired networks happens mostly because of the network congestions packet drop. Therefore data loss is the main signal of congestion and controls the transmission rate in wired networks.  However, although congestion is one of the reasons to lose the packet in wireless networks, interference and signal level may also cause data loss. Hence other techniques are required to adapt wireless networks. Moreover, due to user mobility and channel fadings, network bandwidth and round trip time fluctuations are more common in wireless networks. Then, considering special characteristics of them, they need customized CC techniques to handle their special environments. With regards to network classifications, some of the wireless networks have fixed infrastructure like cellular and wifi, while there are also wireless ad hoc networks where there are no fixed infra structures. Ad hoc networks are also classified into class of mobile ad-hoc network and fixed ad-hoc networks (Figure \ref{fig:secwireless}). In 2001 TCP Westwood \cite{Di2001} was published as a sender-side-only modification to TCP New Reno.

TCP westwood intended to better handle large bandwidth-delay (HBD) product paths, with potential packet loss which is more common in wireless than wired networks. HBD network was discussed before, but a combination of HBD and wireless network is a challenge of heterogeneous networks. Wang et al.  \cite{Wang2016} published the TCP-FIT that performs well over heterogeneous networks that contain both wireless and large BDP links.

In this section we survey different wireless networks and their traffic characteristics, in addition to their challenge in designing efficient transport protocols. In section \cref{sec_cellular} we review the cellular networks, section \cref{sec_sensor_Network} covers fixed ad hoc networks such as sensor networks. Section \cref{sec_Manet_Vanet} discuss Mobile Adhoc Networks (MANET) such as Vehicle Ad Hoc (VANET \cref{sec_vanet})and Flying Ad Hoc (FANET \cref{sec_fanet}) network.

\subsection{CC  in Cellular (3G, 4G, 5G) Networks} \label{sec_cellular}


Fixed bandwidth physical channel is one of the main characteristics of the wired networks. However radio links capacity is very volatile. Radio link speed fluctuates for many reasons such as weather conditions, user to antenna line of sight (which change based on user mobility), and multipath radio fading\cite{Winstein2013,Zaki2015}. Besides in wireless network user handset power is a limiting constraint that needs to be considered to design any new functionality.  Additionally high-resolution live video and Augmented Reality/Virtual Reality are very much growing applications in cellular networks which require high bandwidth and low delay traffic paths. Furthermore cellular networks keep large buffers at base stations to smooth out the bursty data traffic over the time-varying channels and hence bufferbloat is more possible to occur.  Studies in \cite{ guo2016understanding,Huang2013,Jiang2012} confirmed that the bufferbloat problem is becoming a severe challenge in cellular networks. Therefore these networks and their traffic characteristics make them more complex to design an efficient CC mechanism. Multi-gigabit-per-second data rates for fifth generation cellular networks impose even more challenge in designing efficient transport protocol\cite{Zhang2018}.

To alleviate this issue, smartphone vendors put an upper bound on the sender’s congestion window by advertising a static receive window smaller than the actual receive buffer size. Jiang et al. \cite{Jiang2012,Jiang2016} propose the Dynamic Receive Window Adjustment (DRWA) to improve TCP performance over bufferbloat cellular networks. Verus \cite{Zaki2015} is an end-to-end CC protocol that uses delay measurements to adapt the window size.

In continuation with older techniques, some packet marking and rate based TCP algorithms have been proposed to alleviate the buffer bloating problem in cellular networks. Accelerate-Brake Control (ABC) \cite{Goyal2017a} is a mechanism developed to mark each packet with an “accelerate” or “brake” notification as signaling in the base station. This causes the sender to either slightly increase or slightly reduce its congestion window. PropRate \cite{Leong2017} is  a  rate-based TCP algorithm that directly regulates the packets in a bottleneck buffer. This helps to achieve a trade-off in terms of delay and throughput.

As it is discussed before, users' mobility in cellular networks makes highly variable network behaviour. Therefore there is a lot of research on prediction, estimation, forecast, and adaptivity to control the flow in cellular networks \cite{tan2007empirical,ekman2002prediction}. In Sprout \cite{Winstein2013} the receiver makes a short-term forecast of the bottleneck link rate using probabilistic inference. Sprout is a transport protocol for real-time interactive applications over cellular networks. As the primary signal Sprout deploys the receiver’s observed packet arrival times to determine  the quality of the radio link channel at different moments. Then the Sprout receiver sends the forecast to the sender by piggybacking it onto its own outgoing packets. Huang et al. \cite{Huang2013} developed a lightweight passive bandwidth estimation technique for LTE. They show that LTE has lower RTTs than those of 3G networks. ExLL \cite{Park2018} infers the cellular bandwidth through the downlink packet reception pattern. Exll is a low-latency CC that can adapt to dynamic cellular channels without overloading the network, which is mostly popular in probing the dynamics of radio channels. Exll also calibrates the minimum RTT  from the inference on uplink scheduling intervals.

In addition to forecasting the network and user behaviour, cross layer CC is also another approach that researchers have been trying to explore. Consistent congestion and contention cooperation control mechanisms improves the global end-to-end connection and the local wireless link. Therefore authors in  \cite{Chang2017}  propose a Cross-layer-based Adaptive CC (CACC) for the connection-based transport layer and the link-based media access layer. CQIC \cite{Lu2015a} is another cross layer CC mechanism that is proposed in 4G cellular networks. CIQC leverages the physical layer information exchanged between 4G base stations and mobile phones to predict the capacity of an underlying cellular link, and

5G networks provide very high and volatile bandwidth to the users. CC researchers are also studying 5G networks performance and improvement scenarios. Zhang et al. \cite{zhang2016transport} presents a performance evaluation of TCP CC in 5G mmWave cellular. Ford et al. \cite{ford2017achieving} surveys some of the challenges and possible solutions for delivering end-to-end reliable ultra-low-latency services in mmWave cellular systems in terms of the MAC layer, CC, and core network architecture. Network Slicing has been applied as an essential feature of future 5G mobile communication networks. Han et al. \cite{han2018admission} proposes a novel approach of inter-slice admission and CC.  Nasimi et al. \cite{nasimi2018edge} proposed a CC mechanism that functions within the framework of Multi-Access Edge Computing (MEC). They introduced a dedicated function called CC Engine (CCE). CCE captures the Radio Access Network (RAN) condition then utilizes this knowledge to make real time decisions for selectively offloading traffic.

\subsubsection{D2D or M2M }
Device to Device (D2D) or Machine to Machine (M2M) (also known as Internet of Things (IoT)) is another growing application of cellular networks \cite{feng2014device, ghavimi2014m2m,soltanmohammadi2016survey}. M2M communication CC is a key performance challenge to improve.

%
%
%
%

\subsection{CC in Fixed ad hoc networks (Sensor Networks)} \label{sec_sensor_Network}

Challenges in cellular networks CC would be more complicated when there is no infrastructure like base stations. These days a myriad number of sensor network applications are proposed and developed in environmental/earth sensing and industrial monitoring.  Air pollution and water quality monitoring, as well as forest fire and landslide detection are some of the environmental sensing applications of the sensor networks. Machine health monitoring, Water/wastewater, and structural health monitoring are also known as some of the frequent industrial application of the sensor networks.

Sensor network topologies vary based on their applications. It may be very random when a UAV randomly distributes sensors in an inaccessible location. Their topologies may also be neatly designed in advance like industrial applications. Traffic volume in this type of network may not be huge and their applications are very delay sensitive. The constraints in sensor networks are memory, and computational power. In addition to specific network and traffic characteristics, there are performance metrics defined for them specifically.

As it is discussed in \cite{Sergiou2014}, congestion in sensor networks can be detected through buffer occupancy, wireless channel load, and a combination of buffer occupancy. Additionally congestion in sensor networks could be notified explicit implicitly.  A wide variety of the sensor networks CC are surveyed in \cite{wang2006survey,Sergiou2014,Kafi2014,Shah2017,lim2019survey}.

\subsection{CC in Mobile Ad hoc Networks (MANET)} \label{sec_Manet_Vanet}
Vehicular and Flying ad hoc networks are the two recent mobile adhoc systems that are growing rapidly. Smart cities, Internet of Things (IoT), and autonomous driving and flying vehicles are the main reasons for growing MANETs. There are applications running in these networks that generate delay sensitive data traffics. In general CC in MANET can be classified in three classes. 1- packet loss based transport  mechanisms \cite{holland2002analysis, chandran2001feedback,kopparty2002split,liu2001atcp,kim2001tcp,sundaresan2005atp,molia2018tcp,govindarajan2018enhanced}, 2-cross-layer based CC  \cite{hu2004exploiting,sharma2017adaptive,gowtham2019congestion,anuradha2017cross,suraki2018fclcc}, 3-other CC approaches \cite{rajesh2017congestion,vadivel2017adaptive,chen2016joint}. Dimitris Kanellopoulos surveyed these mechanisms in more detail \cite{Kanellopoulos2018}.

\subsubsection{Vehicular Ad-hoc NETworks(VANET)} \label{sec_vanet}
These days vehicles are being equipped with embedded sensors with processing and wireless communication capabilities \cite{Cunha2016}. The VANETs can provide a wide variety of services such as Intelligent Transportation System (ITS) e.g. safety applications. Some applications of VANETs are published such as intersection collision avoidance, information communication with other vehicles, sign extension, vehicle diagnostics, and maintenance. VANET CC protocol design needs to know the network characteristics. VANET characteristics are variable in network density in urban and rural, predictable mobility because of the road maps, no power constraints, rapid changes in network topology, large scale network in dense areas,and  high computational ability \cite{Al-Sultan2014}. VANET mobility is more predictable not only because of the road maps but also because of users' daily behaviour. For example every morning x number of users drive to work in specific repeated paths and the same behavior takes place in the evening. Therefore network traffic density and hourly congestion could be predicted in most cases.\\

\begin{figure}[h]
			\centering
	\pgfdeclarelayer{cars}
	\pgfdeclarelayer{lights}
	\pgfsetlayers{main,lights,cars}
		\resizebox{.5\textwidth}{!}{%

	\begin{tikzpicture}
	\onramproad
	\accident{left}{6}
	
	\commwave{(lane1-left) ++(3,0)}{2}
	\commwave{(lane2-left) ++(4,0)}{2}
	
	\path (lane1-left) +(1,0) node[Car] {};
	
	\tikzstyle{every pin}=[message]
	
	\begin{pgfonlayer}{cars}
	\path (lane1-left) +(4,0) node[Car,pin=below:\textbf{Warning:} emergency message!] (car1) {};
	\path (lane2-left) +(2,0) node[Car,pin=above:\textbf{Warning:} emergency message!] (car2) {};
	\end{pgfonlayer}
	
	\begin{pgfonlayer}{lights}
	\foreach \i in {1,2} {%
		\fill[red] (car\i.north west) +(4pt,-4pt) circle (3pt);
		\fill[red] (car\i.south west) +(2pt,4pt) circle (3pt);
	}
	\end{pgfonlayer}
	
	\end{tikzpicture}
}
	\caption{Accident warning message broadcast back to approaching vehicles in VANET.}
	\label{fig:VANET}
\end{figure}
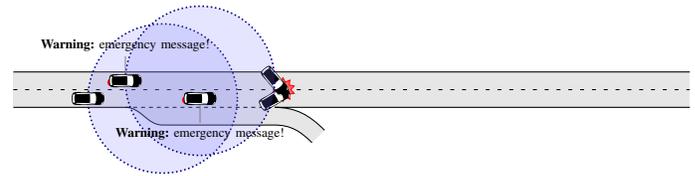

Safety message scheduling and distributed CC are the two types of studies that are growing in VANETs (Figure \ref{fig:VANET}). DRCV \cite{drigo2009distributed} monitors and estimates channel load and controls the packet rate of outgoing periodic when event-driven safety packets are detected, and the rate of periodic packets has to be dropped promptly therefore Fast drop mechanism is developed to manage this part. MOTabu \cite{Taherkhani2015} is a dynamic and distributed mechanism. It consists of two components: congestion detection and CC. Channel usage level is used to detect and it is controlled by  MOTabu algorithm that is used to tune transmission range and rate for both safety and non-safety massages by minimizing delay and jitter. Gatekeeper controls the packet rate as a function of the channel load  in European standardization. Kuhlmorgen et  al. \cite{kuhlmorgen2017evaluation} studied the performance of the gatekeeper with packet prioritization and an adaptive linear control algorithm. This is designed to make sure that high vehicle density and channel load conditions do not impact the efficiency and safety of the applications.  DySch and TaSch  \cite{taherkhani2016prioritizing} are strategies to assign priorities to the safety and service messages. These priorities are defined based on the content of messages as a static factor, state of network as a dynamic factor and size of messages. This technique helps to increase reliability and safety by giving higher priority to the safety messages. Kuhlmorgen et al. \cite{kuhlmorgen2016impact} evaluate the performance of contention-based forwarding (CBF) under congestion-free and saturated channel conditions for several protocol variants. Their results show that the standard distributed CC (DCC)-gatekeeper hinders the functionality of CBF even for small node densities. They also proposed an enhanced  DCC-gatekeeper-based approach.

Reliability of the safety applications is one of the main concerns on designing VANET. Network congestion is one of the main reasons behind unreliable communication. Therefore CC for reliable safety communication is a research topic that researchers are trying to improve.  Beacon and event-driven warning are two important messages that have been utilized to improve road safety and traffic efficiency in VANET.  Beacons are periodically broadcasted by vehicles to inform their neighbors of their information. Le et al.  \cite{le2011performance} considered three beacon CC algorithms that control the beacon load below a certain threshold by adjusting the transmit rate and/or transmit power for the beacon messages.  Darus et al. \cite{darus2013congestion} developed a CC algorithm to provide reliable event- driven safety messages. Bouassida et al. \cite{ bouassida2008congestion} presented the concept of dynamic priorities-based scheduling, to ensure a reliable and safe communications architecture. Messages priorities are dynamically evaluated according to their types, the network context and the neighborhood. Qureshi et al. \cite{qureshi2018dynamic} surveyed the CC of VANET and they proposed a Dynamic CC Scheme (DCCS) as a means of reliable and timely data delivery, in safety applications.  An adaptive beacon generation rate (ABGR) CC mechanism is also proposed \cite{li2020reliable} to reduce channel congestion and contention. A survey of CC mechanisms on VANET is published in \cite{Taleb2017}. Named data networking is a data oriented communication model that is also applied on VANET. Author's in  \cite{Shemsi2017,khelifi2019named} surveyed named data networking in VANET.

\subsubsection{ Flying Ad hoc Network (FANET)} \label{sec_fanet}
Applications of swarming Unmanned Aerial Vehicle (UAV) are growing in smart cities. UAVs communication is one of the key factors to improve their flight performance and security. Therefore Flying Ad Hoc Networks and their transport protocols are a new research area. Authors in \cite{bekmezci2013flying,khan2017flying,aruna2019adaptive} reviewed the FANET communication, architecture and routing protocols.

\section{Latency and Throughput Optimization (L\&T) in Application Layer } \label{sec_app_cc}
Application layer transport control is another method that is used in some technologies. Constrained Application Protocol (CoAP) used for Internet of Things (IoT) devices. IoT will be more discussed in  \cref{sec_IoT}. Quic is also  another application layer transport protocol developed by Google. Quic is covered in  \cref{sec_quic} .

\subsection{Internet of Things (IoT), CoAP} \label{sec_IoT}
The Internet of Things is also a recent technology that is growing very fast. Many devices such as sensors and actuators communicate together to form an IoT system. \cite{gubbi2013internet} extensively discuss the IoT, vision, architectural elements, and future directions. IoT Applications, investments, and challenges for enterprises are discussed in \cite{lee2015internet}, and Farooq et al. reviews the IoT in \cite{ farooq2015review}. It is different from the previous networks because there is less control over the middle way nodes to control the congestion. Besides, the traffic pattern of this system is different from previous networks. IoT devices are also power constrained devices, hence CC protocol should be designed in a way to consume the minimum possible power. Buffer overflow happens frequently in this technology since IoT devices utilize shallow buffers and they have lossy transmissions mostly because of the wireless network.

The  Extensible Messaging and Presence Protocol (XMPP), Message Queuing Telemetry Transport (MQTT), and Advanced Message Queuing Protocol (AMQP)  \cite{saint2004extensible,verma2020iot}  are IoT application protocols which use the TCP  as a transport layer protocol to offer data transmission. However Constrained Application Protocol CoAP  is another IoT application protocol designed to provide a fast connection between devices. CoAP runs over the User Datagram Protocol (UDP) \cite{verma2020iot}. CoAP is a lightweight RESTful application layer protocol that is developed for the IoT. CoAP must handle congestion control by itself since CoAP operates on top of UDP. CoAP utilizes the confirmable or non-confirmable messages to communicate. Default CoAP applies random initial retransmission time outl (RTO), doubling the RTO value of the retransmitted packet if the ACK is not received (loss). In response to detecting packet loss, doubling the RTO also called Binary Exponential Backoff (BEB) is applied to overcome the congested communication in CoAP\cite{bormann2014coap,betzler2016coap}.

RTO and backoff estimation are the topics that researchers try to improve. \cite{balandina2013computing}  is proposed to find RTO based on the ratio between current sample of the round-trip time (RTT) and the RTO value. Later on a 4-state estimator for variable backoffs is proposed in CoCoA  \cite{Bhalerao2016} to a balance between maximizing throughput and minimizing packet loss. CoCoA+ \cite{Betzler2015} improves the performance of CoCoA mechanism in three stages. 1- initial RTT estimator is modified.  2- BEB is replaced for retransmissions by a Variable Backoff Factor (VBF). 3- aging approach for large RTO. Bolettieri et al. \cite{Bolettieri2018} proposed a RTO estimation modification as precise Congestion Control Algorithm (pCoCoA). Different congestion control mechanisms are surveyed in \cite{maheshwari2020analysis}.

CoRE is an alternative to CoAP congestion control approach that is a topic of interest in the IETF Working Group.  Betzler et al. \cite{betzler2013congestion} evaluate and show how the default and alternative congestion control mechanisms compare to each other.

\subsection{QUIC} \label{sec_quic}
HTTP/2 is developed by Google to run over TCP to multiplexe multiple connection stream. It fixes the head-of-line (HoL) blocking problem.  QUIC establishes a number of multiplexed connections  over User Datagram Protocol (UDP) to resolve the HoL problem. QUIC is also designed to reduce the connection and transport latency, and bandwidth estimation in each direction to avoid congestion \cite{gratzer2016quic,Langley2017,Cui2017}.  Additionally QUIC  moves the congestion control algorithms into the user space at both endpoints, rather than the kernel space.

Research studies are focused on comparing the performance of QUIC and other alternatives while they are running on different networks. Megyesi et al. \cite{Megyesi2016} studied the performance of QUIC, SPDY (precedent of QUIC) and HTTP. They found that their performance depends on actual network conditions. Carlucci et al. \cite{Carlucci2015} ran the same evaluation. They have found that QUIC reduces the overall page retrieval time with respect to HTTP in case of a channel without induced random losses and outperforms SPDY in the case of a lossy channel. Kakhki et al. \cite{Kakhki2017} identified performance issues related to window sizes, reordered packets, and multiplexing large numbers of small objects. Further they identify that QUIC’s performance diminishes on mobile devices and over cellular networks. Biswal et al.  \cite{Biswal2016} found that QUIC performs better under poor network conditions such as low bandwidth, high latency, and high loss. Manzoor et al. \cite{manzoor2019improving} showed that QUIC achieves sub-optimal throughput in WiFi networks. They tuned QUIC to produce bursty traffic called Bursty QUIC (BQUIC) . BQUIC achieves better performance in WiFi. In addition to the delay, security of the QUIC is a key factor to consider. Saverimoutou et al. \cite{Saverimoutou2017} studied the security aspects of the TLS/TCP and QUIC/UDP. They identify some vulnerabilities of two protocols and evaluate their impacts on HTTP/2-based web services.


In current congestion control mechanism, data paths are becoming diverse such as Linux kernel TCP, UDP-based QUIC, or kernel-bypass transports like mTCP-on-DPDK \cite{jeong2014mtcp}. Therefore separating the congestion control agent outside of the datapath is another interesting research topic \cite{Narayan2017,narayan2018restructuring} called Congestion Control Plane (CCP). CCP improves both the pace of development and improve the maintenance of congestion control algorithms. Kaufmann et al.  in \cite{kaufmann2019tas} presents TAS, which is an implementation of moving congestion control out of the datapath to a dedicated CPU.

\section{Latency and Throughput Optimization (L\&T) through Machine Learning/Neural  based Algorithms}\label{sec_ML}
 Machine Leanrning (ML) algorithms have been growing exponentially over the last decade. They improved the performance of different complex problems of various fields significatly.  
ML algorithms are evolved in three main groups of prediction, description and control problems. 

\subsection{ML introduction} \label{sec_MLintro}
There are three main groups of machine learning algorithms that are growing exponentially. Predictive, descriptive and Reinforcement Learning controllers.

\paragraph{Predictive}Prediction algorithms try to forcast the future events based on previuous labeled information as training datasets. These algorithms are also called supervised learning or in some other references, they are called generalization and function estimation. Regression and classifications are the two most common works on prediction problems. ML algorithms could be also classified as parametric and non parametric.

 Paramettric algorithms work as function estimation and optimizing the parameters of assumed functions. Linear and logistic regression work as function estimation and binary classifier respectively. Linear regression is a linear function of the input predictors  $y= \beta_0+\beta_1x_1+...+\beta_nx_n$, where n is number of predictors. Minimizing the error between input data and estimator helps to find the function parameters. Logistic regression assumes a sigmoid function ($y=\frac{e^x}{1+e^x}$) as relation between predictors and the output. Different error (cost) optimization such as gradient descent can be applied to find the function parameters.

Non-parameteric algorithms do not make strong assumption about the form of mapping functions, therefore they are more flexible to fit training data.
Decision tree and it's derivatives such as random forest, gboost, and xgboost are  a group of well-knwon non-parameteric predictors. k- Nearest Neighbors  (kNN) and Support Vector Machine  (SVM) are the other two non-parameteric supervised ML algorithms that are applied on various of problems.

\paragraph{Descriptive}descriptive machine learning algorithms are also known as clustering or unsupervised learning. These algorithms try to cluster a set of objects in such a way that objects in the same group  are more similar (in some sense) to each other than to those in other groups. Density based \cite{erman2006traffic} and hierarchical clustering \cite{ma2006unexpected} are the two lagorithm that are used in network traffic clustering.

\paragraph{Reinforcement Learning, Controller} 
RL algorithms are a subset of ML that interact with environment to train and then deploy as controller. Focus of the RL algorithms is on finding a balance between exploration (of uncharted territory) and exploitation (of current knowledge. RL agents learn to choose actions that maximize the cumulative rewards based on the environment state. RL uses the Markov Decision Process (MDP) to represent the environment. Dynamic programming \cite{bellman1966dynamic,bertsekas1995dynamic,puterman2014markov} is the most common technique used to solve MDPs. Q-learning \cite{watkins1992q} is another approach for a model-free reinforcement learning algorithm that is developed to solve the MDP problems with stochastic state transitions and rewards. Figure \ref{fig:RL}. represents the RL diagram. Network flow rate control can be also seen as a sequence of decisions to increase or decrease the transmission rate. This flow rate decision helps transmitters to adapt the network load and behavior. Environment states and rewarded feedback are the two main components that RL algorithms need as input to work. Then RL agents train to generate actions according to the environment state.

\tikzstyle{block} = [rectangle, draw, 
text width=8em, text centered, rounded corners, minimum height=4em]

\tikzstyle{line} = [draw, -latex]
\begin{figure} [h]
	\centering
	\resizebox{.5\textwidth}{!}{
		\begin{tikzpicture}[node distance = 6em, auto, thick]
		\node [block] (Agent) {Agent};
		\node [block, below of=Agent] (Environment) {Environment};
		
		\path [line] (Agent.0) --++ (4em,0em) |- node [near start]{Action $a_t$} (Environment.0);
		\path [line] (Environment.190) --++ (-6em,0em) |- node [near start] {New state  $s_{t}$} (Agent.170);
		\path [line] (Environment.170) --++ (-4.25em,0em) |- node [near start, right] {Reward $r_{t}$} (Agent.190);
		\end{tikzpicture}
	}
	\caption{Reinforcement learning, Agent and Environment interaction diagram.} \label{fig:rldiagram}
	\label{fig:RL}
\end{figure}
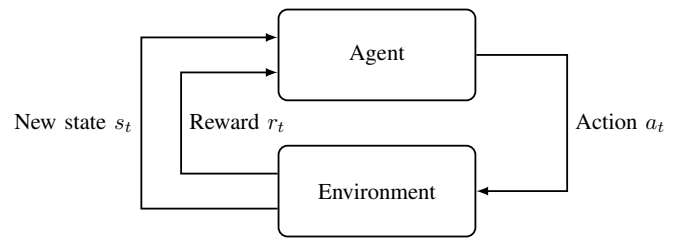

In 2015 authors in \cite{mnih2015human} inspired from psychological and neuroscientific perspectives of animals  integrate the deep learning \cite{lecun2015deep,goodfellow2016deep} and RL to develop the idea of Deep Reinforcement Learning (DRL) (Figure \ref{fig:drlDiagram}). A few months later, leveraging the DRL, Alphago and other atari games \cite{silver2016mastering,silver2017mastering,mnih2013playing} were the computer program to defeat world champion challenge. This shows that DRL are able to achieve beyond human capability. This game is known to be googol times more complex than chess.  Humanoid is another substantial presentation of the DRL capabilities. DRL is also applied on other games to investigate a variety of decision sequences and DRL performance \cite{shao2019survey}. These days Deep Reinforcement Learning is significantly popular because of its glorious result on different control and decision making problems.

Integrating deep learning in RL agents, facilitates the increasing size of environment states dimensions. Some of the most popular recent continuous states DRL algorithms are deep reinforcement learning with double Q-learning \cite{van2016deep}, Deep Deterministic Policy Gradient (DDPG) \cite{lillicrap2015continuous},  Luong et al. surveyed deep reinforcement learning applications in communications and networking \cite{luong2019applications}.

Some DRL applications on networks are limited to discrete and finite actions and states. Mao et al. \cite{mao2016resource} proposed a network resource management that is developed based on finite actions DRL algorithms. 
TCP-Drinc \cite{xiao2019tcp} is a deep reinforcement learning based congestion control that uses a finite set of actions to resize the TCP window in five different ways : $A=\{w=w \pm 1 ; w=w\pm \frac{1}{w} ; no\: change\}$.  Experience-driven networking \cite{xu2018experience} is another DRL based approach for traffic engineering problems. DRL-TE uses a finite set of throughput and delay as discrete state space as well as a finite set of actions. It's reward is defined as the total utility of all the communication sessions. \\

\tikzset{%
	neuron missing/.style={
		draw=none, 
		scale=2,
		text height=0.333cm,
		execute at begin node=\color{black}$\vdots$
	},
}

\newcommand{\DrawNeuronalNetwork}[2][]{
	\xdef\Xmax{0}
	\foreach \Layer/\X/\Col/\Miss/\Lab/\Count/\Content [count=\Y] in {#2}
	{\pgfmathsetmacro{\Xmax}{max(\X,\Xmax)}
		\xdef\Xmax{\Xmax}
		\xdef\Ymax{\Y}
	}
	\foreach \Layer/\X/\Col/\Miss/\Lab/\Count/\Content [count=\Y] in {#2}
	{\node[anchor=south] at ({2*\Y},{\Xmax/2+0.1}) {\Layer};
		\foreach \m in {1,...,\X}
		{
			\ifnum\m=\Miss
			\node [neuron missing] (neuron-\Y-\m) at ({2*\Y},{\X/2-\m}) {};
			\else
			\node [circle,fill=\Col!50,minimum size=1cm] (neuron-\Y-\m) at 
			({2*\Y},{\X/2-\m}) {\Content};
			\ifnum\Y=1
			\else
			\pgfmathtruncatemacro{\LastY}{\Y-1}
			\foreach \Z in {1,...,\LastX}
			{
				\ifnum\Z=\LastMiss
				\else
				\draw[->] (neuron-\LastY-\Z) -- (neuron-\Y-\m);
				\fi
			}
			\fi
			\fi
			\ifnum\Y=1
			\ifnum\m=\X
			\draw [overlay] (neuron-\Y-\m) -- (state);
			\else
			\ifnum\m=\Miss
			\else
			\draw [overlay] (neuron-\Y-\m) -- (state);
			\fi
			\fi
			\else
			\fi     
		}
		\xdef\LastMiss{\Miss}
		\xdef\LastX{\X}
	}
}
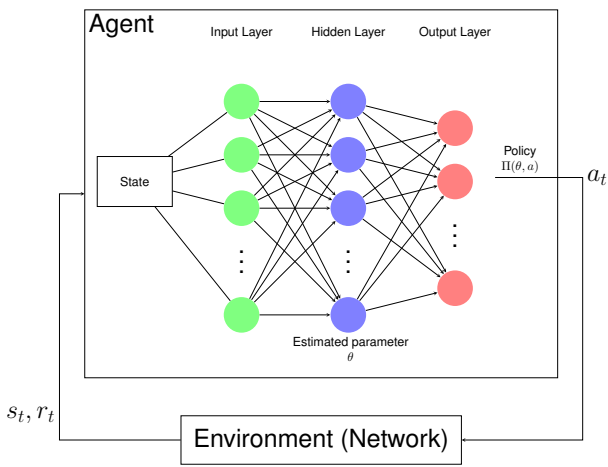
\begin{figure}[h]
		\centering
	\resizebox{.45\textwidth}{!}{%

		\begin{tikzpicture}[x=1.5cm, y=1.5cm,
		>=stealth,font=\sffamily,nodes={align=center}]
		\begin{scope}[local bounding box=T]
		\path  node[draw,minimum width=6em,minimum height=4em] (state) {State};
		\begin{scope}[local bounding box=NN]
		\DrawNeuronalNetwork{Input Layer/5/green/4///,
			Hidden Layer/5/blue/4//11/,
			Output Layer/4/red/3//11/}
		\end{scope}
		\path (NN.south) node[below]{Estimated parameter\\ $\theta$};
		\path(NN.east) -- node[above]{Policy\\ $\Pi(\theta,a)$}++ (4em,0);
		\end{scope} 
		\node[fit=(T),label={[anchor=north west]north west:\huge Agent},inner sep=1em,draw]
		(TF){};
		\node[below=3em of TF,draw,inner sep=1em] (Env) {\huge Environment (Network)};
		\draw[<-] (TF.180) -- ++ (-2em,0) |- (Env.180) node[pos=0.45,left]{\huge $s_t, r_t$};
		\draw[->] (NN.east) -- ++ (7em,0)node[right]{\huge $a_t$} |- (Env);
		\end{tikzpicture}
	}
	\caption{Deep Reinforcement Learning model.}
	\label{fig:drlDiagram}
\end{figure}

Utility function extraction is also a technique that is tried in different studies. In this method researchers try to find out the network utility as a function of feedback signals and transmission rate. Utility function later extends to reinforcement learning algorithms to control the network flows. A utility-based congestion control scheme for Internet-style networks with delay is presented in  \cite{alpcan2003utility}.  In \cite{Winstein13book} the state of the network is defined in three variables as: 1- An exponentially-weighted moving average (EWMA) of the interarrival time between new acknowledgments received (ack\_ewma), 2- An exponentially-weighted moving average of the time between TCP sender timestamps reflected in those acknowledg- ments (send\_ewma), and 3- The ratio between the most recent RTT and the minimum RTT seen during the current connection (rtt\_ratio). By learning the dynamics of the network environment the DQL agent aims to maximize the total utility of all the communication sessions, which is defined based on end-to-end throughput and delay. They use two utility objective functions as  $U = log(throughput) - \delta . log(delay)$ and  $U=-\frac{1}{ throughput}$. Increasing throughput or decreasing delay increases the first utility function while only increasing the throughput can increase the second utility function. They achieve better results in cellular network situations, which could have been expected since utility objective function is tracking the throughput fluctuation in cellular networks. In the case of data centers, there is no improvement in their results since throughput in data centers are not much fluctuating like cellular networks.

Performance-oriented Congestion Control (PCC)  \cite{Donga2015} is another attempt at designing a utility model. They defined the utility function as $U(x) = T.Sigmoid_{\alpha} (L  - 0.05) - x. L$. where x is the sending rate, L is the observed data loss rate, T = x(1 - L) is sender's  throughput, and  $Sigmoid_{\alpha} (y)= \frac{1}{1+e^{\alpha y}} $ for some $\alpha >0$. PCC Vivace \cite{Dong2018vivace} is designed within the PCC framework to leverage machine learning based optimization. Mobile Oriented Rate Control (MORC) \cite{Mangiante2018} builds based on the PCC protocol design framework to achieve high throughput and low latency. MORC uses the achieved goodput, packet loss rate, and average latency to build utility functions and then uses an online learning optimization algorithm to select the best rate.

Verus \cite{Zaki2015} is another attempt to extract a network delay profil to apply on TCP window size. The key idea of Verus is to continuously learn a delay profile that captures the relationship between end-to-end packet delay and maximum window size over a short monitoring interval, and uses this relationship to increment or decrement the window size based on the observed short-term packet delay variations. At each time instance,Verus either: increments or decrements W, using the delay profile as W(t + 1) = f(d(t) + $\delta (t))$ , where W (t + 1) is the next sending window, f is the delay profile function with d(t) being the network delay, and $\delta (t)$ is a delay increment or decrement value. Copa \cite{arun2018copa} is a  method that presents a target rate equals a function of throughput and delay ($\frac{1}{\delta . d_q}$) under a Markovian packet arrival model (steady state). $\delta$ is a weighting for delay and throughput.\\


Machine learning algorithms are applied in different data networks' problems. Authors in \cite{wang2017machine,boutaba2018comprehensive,polese2019survey,cheng2019bridging,sun2019application} surveyed various machine learning approaches to resolve a variety of the network problems. In terms of delay and throughput improvmenet, Predictive and RL controller ML algorithms have been applied on problems such as  queue management, packet loss classification, congestion window update, congestion inference, and admission control, resource allocation.	

\subsection{Queue Management}\label{sec_ML_QManagement}
Managing the buffer queues in intermediate nodes (switches/router) have been one of the complicated problems.Conventional queue management have problem to utilize links efficiently, queue delays and unfairness. Advanced Active Queue Management (AQM) schemes also suffer from poor responsiveness,  queue length stabiliity  problem, and its performance (utilization and drop) highly depends on parameter tuning. Many different approches have been published either as predictors or controllers. Gao et al. \cite{gao2002exploiting} designed predictive AQM as a generalized version of RED with traffic prediction. Their approach stabilize the queue length at a desirable level and enables the link capacity to be fully utilized, while not incurring excessive packet loss. 
DEEP BLUE \cite{masoumzadeh2009deep} is proposed  as DRL based AQM scheme. Kim et al. \cite{kim2019deep} investigate the Q-learning DRL algorithm to manage the IoT network queues.

\subsection{Packet Loss classification} 	\label{sec_ML_LossClasification}
Packet loss classification is a key feature in wireless network because loss due to radio link fading problem could be distinguished from congestion based loss. Authors in \cite{liu2003end,barman2004model,el2005improving,el2010enhancement} proposed different algorithms to classify the packet losses.
\subsection{Congestion Control} 	\label{sec_ML_CC}
Network state predictions are widely applied in different networks to proactively control the data flow. Sprout \cite{Winstein2013} is a transport protocol for real-time interactive applications over paths that traverse the cellular wireless networks. Sprout uses packet arrival times as a congestion signal. Probabilistic inference is also used to forecast the packet deliveries. Sender uses this forecast to pace its transmissions. Random forest \cite{yue2017linkforecast} is another machine learning technique that is used to forecast the volatile bandwidth in LTE cellular networks. Other ML algorithms are also tried in network link congestion prediction. Wu et al. in  \cite{wu2019link} presented the Support Vector Machine (SVM)\cite{tian2012recent}, Multi-Layer Perceptron (MLP) \cite{nikravesh2016mobile}, Onedimensional Convolutional Neural Network \cite{kim2014convolutional,krizhevsky2012imagenet}(1DCNN) and K-Nearest Neighbor (KNN) \cite{cover1967nearest}, to predict the network link congestion for Software-Defined-Network Data Plane.

Iroko  \cite{ruffy2018iroko} is designed as a framework to prototype reinforcement learning for data center traffic control. Iroko uses the DDPG and PPO algorithm to train the DRL agent. Iroko uses the switch buffer occupancy, interface utilization, and active flows as the network (environment) states. Actions are defined as  $ bw_i \leftarrow bw_{\max} * a_i \forall i \in hosts$ and reward function is :

\[
R \leftarrow \sum_{i \in hosts}
\underbrace{\frac{bw_i}{bw_{max}}}_\text{bandwidth reward} - 
\underbrace{ifaces}_\text{weight} * \underbrace{(\frac{q_i}{q_{max}})^2}_\text{queue penalty}  -\underbrace{std}_\text{devpenalty}  
\]
 
DRL is also applied to control the congestion in cellular networks \cite{nascimento2019deep}. In this study DQN  \cite{mnih2013playing} and A3C \cite{mnih2016asynchronous} are used to train the agent. Observation state space defined as the EWMA of 1- the time interval between two sent packets, 2- the time interval between two consecutive ACKs, and 3- the RTT (round trip time). The action space is a set of three possible decisions to be enforced over the TCP congestion window  (cwnd): 1- cwnd incremented (50 bytes), 2- cwnd decremented (-10 bytes) and, 3- cwnd not changed. The reward function was defined based on the throughput per flow, being calculated as: throughput = (bits/latency(s))

Cui et al. \cite{cui2020improving} proposes Hd-TCP as a DRL based Congestion Control  for mobile networks in High-Speed Railway (HSR). Hd-TCP is designed to deal with frequent handover on HSR from the transport layer perspective. Hd-TCP uses state vector is 1) goodput: an Exponentially-Weighted Moving Average (EWMA) of the receiving rate, defined as the number of ACK’ed bytes during an RTT, divided by the RTT,  2) cwnd: the current congestion window size, 3) avgRTT: the average RTT during a time step, 4) queue-delay (dq): the queuing delay during a time step, measured as the difference between the avgRTT in the same time step and the estimated RTprop (minimum RTT) in the time window of 10 s, 5) loss-count: the number of lost packets during a time step,and  6) rsrp: the reference signal receives power, the signal power between UE and base station, measured from the LTE physical layer information perceived by the UE. Actions and rewards in Hd-TCP are defined as Table \ref{tab:HD-TCP-ACTIONS} and Equation \ref{eq:HD_TCP_reward}. Actions are discrete and finite in this study.

\begin{table}[h]
	\centering
	\begin{tabular}{lc}
		\multicolumn{1}{c}{Actions}          & Change size(Byte) \\ \hline
		\multicolumn{1}{l|}{Rapid Increase}  & 520               \\ \hline
		\multicolumn{1}{l|}{Increase}        & 50                \\ \hline
		\multicolumn{1}{l|}{Rapid Dencrease} & -520              \\ \hline
		\multicolumn{1}{l|}{Decrease}       & -50               \\ \hline
		\multicolumn{1}{l|}{No Change}       & 0                 \\ \hline
	\end{tabular}
	\caption{ACTIONS OF HD-TCP, \cite{cui2020improving}}
	\label{tab:HD-TCP-ACTIONS}
\end{table}

\begin{equation}
reward =
\begin{cases}
log(throughput) - log(dq), & \text{if $dq \le \epsilon$}\\
log(throughput) - 4 * log(dq), & \text{if $dq > \epsilon$}
\end{cases}      
\label{eq:HD_TCP_reward} 
\end{equation}

More recent DRL training algorithms are more efficient, therefore continuous action and state space are possible to process. This continuity improves the fine control problem. Asynchronous Advantage Actor-Critic (A3C) \cite{mnih2016asynchronous}, Trust Region Policy Optimization (TRPO) \cite{schulman2015trust}, and proximal policy optimization (PPO) \cite{schulman2017proximal} are recent approaches that enable the DRL continuous actions. Multiple agent DRL is also another approach to control multiple related problems.  DRL-CC \cite{xu2019experience} is proposed to dynamically and jointly perform congestion control for all active MPTCP flows on an end host. DRL-CC utilizes a single  agent instead of multiple independent agents. The objective of  trained agents is to maximize the overall utility. They applied the LSTM-based representation network into an actor-critic framework for continuous (congestion) control. In this paper state is represented as an array of sending rate, goodput, average RTT, the mean deviation of RTTs, and the congestion window size. These variables are considered for each subflows separately.  Action is defined as an output array that each element corresponding to change needs to be made to the congestion window of that subflow.  Reward is the sum of the utility of all subflows that is defined as different functions of throughput,  delay, $\alpha$-fairness \cite{srikant2004mathematics}.\\

DRL is also applied on NDN \cite{lan2019deep} as emerging future network architecture.  Network state and actions are represented by Table \ref{tab:NDNDRLState} and \ref{tab:NDNDRLActions} consecutively. Reward function is also defined as Equation \ref{eq:rewardDRLNDN}:

\begin{equation}
\begin{split}
Utility_i(t) = \alpha_i.log(throughput_i(t))-\beta_i.RTT_i(t)-\\
\gamma_i.loss_i(t)-\delta_i.reordering_i(t)
\end{split}
\label{eq:rewardDRLNDN}
\end{equation}

Where  $i \in {1, 2, ..., N }$ and $t > 0$ represent different types of contents and different monitor intervals. $\alpha_i , \beta_i , \gamma_i$ and $\delta_i$ are the parameters which control the relative weight or importance of throughput, average RTT, loss rate and packet reordering in the current monitor interval. They applied Deep Q-learning (DQN) to train the agent. They use seven hidden layers in their agent's neural network. Their neural network consists of ten inputs, seven outputs and 128 nodes per each hidden layer.

\begin{table}[h]
	\centering
	\resizebox{.45\textwidth}{!}{
		\begin{tabular}{ll}
			\hline
			Variable    & Definition                                                        \\ \hline
			i\_prefix   & The prefix of requesting content                                  \\
			i\_priority & The priority of parameter of requesting content                   \\
			i\_cwnd     & The immediate window of sending interest packets                  \\
			i\_count    & Total number of interest packets sent during the monitor interval \\
			d\_count    & Total number of Data packets received during the monitor interval \\
			l\_count    & Total number of packets retransmitted during the monitor interval \\
			d\_size     & Average size of Data packets received during the monitor interval \\
			d\_rtt      & Average RTT of Data packets received during the monitor interval  \\
			m\_time     & The time of monitor interval                                      \\
			d\_time     & The time of decision intervals                                    \\ \hline
			&                                                                  
		\end{tabular}
	}
	\caption{ NDN State \cite{lan2019deep}.}
	\label{tab:NDNDRLState}
\end{table}

\begin{table}[h]
	\centering
	\begin{tabular}{ll}
		\hline
		Way of change & Extend of changes (condo) \\ \hline
		Increase      & +1, * 1.25, *1.5          \\
		Decrease      & -1, * 0.75, *0.5          \\
		Maintain      & 0                         \\ \hline
	\end{tabular}
	\caption{ NDN Actions \cite{lan2019deep}.}
	\label{tab:NDNDRLActions}
\end{table}

\subsection{Resource Allocation} 	\label{sec_ML_resourceAllocation}
Predicting demand variability and future resource utilization is the main reason to consider predictive ML algorithms \cite{mijumbi2014design}. 
Network Functions Virtualisation (NFV) is growing to enable service providers to offer more flexible software-defined network functions. Mijumbi et al. \cite{mijumbi2016connectionist} proposed a Neural Network algorithm to predict future required resources on NFV.  Shi et al. \cite{shi2015mdp}  applied MDP to dynamically allocate cloud resources for NFV components. 
Testolin et al. \cite{testolin2014machine}  propose a machine learning approach to support QoE-based Video Admission Control (VAC) and Resource Management (RM) algorithms.

\subsection{Admission Control} 	\label{sec_ML_Admission_control}
Cellular wireless networks suffer from radio link resource limitation more than the other networks. Besides, users mobility and handover make this problem even worse.  Therefore admission control is one of the key points to achieve highest performance and resource utilization.
Bojovi et al. \cite{bojovic2012cognitive}  propose a multilayer feed-forward Neural Network (NN) radio admission control for LTE networks. Their radio admission control  learns from the past experience how the admission of a new session would affect the QoS of all sessions in the future.
Wang et al. \cite{wang2013new} introduce a call admission control (CAC)  for LTE femtocell network.  This the multi-service CAC strategy composes of two parts: subscriber authentication and queuing admission control.
\subsection{Congestion Window} 	\label{sec_ML_Congestion Window}

Reinforcement learning controllers have been applied to many different decision sequence control problems such as network and flow control.  Q-learning based RL controller is applied to design a TCP algorithm for memory constrained IoTs  \cite{li2016learning}. QTCP \cite{li2018qtcp} is another approach to control the TCP flow through Q-learning reinforcement learning. As the network status, QTCP uses its statistics such as: 1-avg\_send: the average interval between sending two packets., 2- avg\_ack: the average interval between receiving two consecutive ACKs, and 3- avg\_rtt: the average RTT.  Three actions are designed in QTCP as Increase (10byte), decrease (-1byte) and no change. They tend to encourage the agent to quickly increase. In  addition, QTCP uses the difference between consecutive utility values as a reward function (Equation \ref{eq:utility_qttcp}). QTCP uses the Q- learning algorithm to train the agent.

\begin{equation}
Utility = \alpha \times log(throughput) - \delta \times log(RTT)
\label{eq:utility_qttcp}
\end{equation}
where $\alpha$ and $\delta$ are two adjustable weights for throughput and RTT. 
Kong et al. designed and implemented RL-TCP \cite{kong_improving_2018}. They take the network state as  an array of EWMA of the ACK interarrival time, EWMA of the packet inter-sending time, the ratio of current RTT and the minimum RTT, the slow start threshold, and cwnd size. They design reward function as the changes of utility U (Equation \ref{eq:utility_RL-TCP}). The utility U is a function of throughput $t_p$, delay $d = RTT - RTT_{min}$, and packet loss rate p. Besides, discreet action is chosen as cwnd = cwnd + x, where x = -1, 0, +1, +3. They applied SARSA \cite{sutton2018reinforcement} as RL algorithms.

\begin{equation}
U=log(\frac{tp}{B})-\delta_1.log(d) + \delta_2.log(1-p)
\label{eq:utility_RL-TCP}
\end{equation}
where $\delta_1$ and $\delta_2$ are two adjustable weights for delay and packet loss rate. Authors in \cite{hwang2005reinforcement,shaio2005reinforcement} proposed other different reinforcement learning approaches to control the congestion of high-speed multimedia networks.

Jay et al. \cite{jay2019deep} applied DRL in Internet congestion control problem to distinguish non-congestion loss from congestion induced loss. Network state is defined as an array of three network statistics and their k step history : 1- latency gradient \cite{Dong2018vivace}, the derivative of latency with respect to time; 2- latency ratio \cite{Winstein13book}, the ratio of the current MI’s mean latency to minimum observed mean latency of any MI in the connection’s history; and 3- sending ratio, the ratio of packets sent to packets acknowledged by the receiver. In this study, reward is defined as $10 \times throughput - 1000 \times latency - 2000 \times  loss$. Action is selected as a real continuous value to fine control the rate and PPO \cite{schulman2017proximal} is used to train the agent.

TCP initial rate control is also a problem that DRL researchers are working to improve. Nie et al. propose  \cite{nie2019dynamic}  dynamic TCP initial windows and congestion control  through reinforcement learning. They utilize A3C for the congestion control and input states are $s_t = (Throughput_t, \: RTT_t,\: Loss_t)$, action is continuous space and reward is defined as  $r_t = log(\frac{Throughput_t}{RTT_t})$


\section {Lessons Learned and Research Opportunities} \label{sec_conclusion}
Although network traffic management and control have been substantially evolved to improved the latency and throughput of different networks,
, our studies and many others in the literature indicate that there is no existing single mechanism that optimally works in all situations. Besides, the set of situations are rapidly increasing. Therefore maybe machine learning based mechanisms would increase to cover different situations. Recent DRL algorithms such as imitation learning and meta learning would have more chance to develop in network flow, congestion control and traffic management. 


Additionally, more flexible and high speed programmable networks would be growing.  
\ifCLASSOPTIONcaptionsoff
  \newpage
\fi



\bibliographystyle{IEEEtran}

\bibliography{output}
%

%
%

%

\begin{IEEEbiography}
	[{\includegraphics[width=1in,height=1.25in,clip,keepaspectratio]{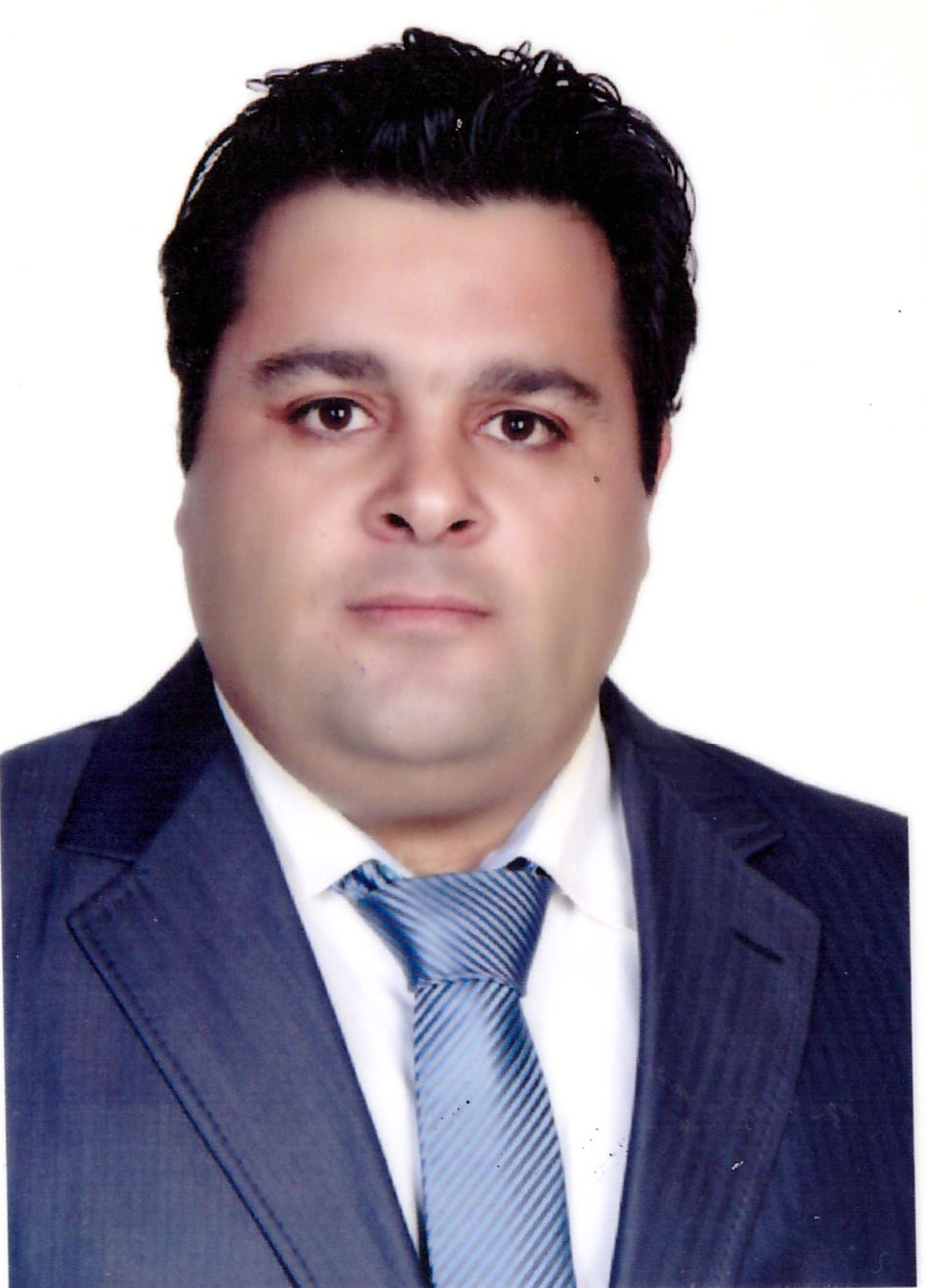}}]{Amir Mirzaeinia}

Received his Masters degree in Telecommunications from Isfahan University of Technology. Amir Mirzaeinia is currently a Computer Science PhD candidate in New Mexico institute of Mining and Technology. Amir Mirzaeinia had been working with Mobile Commmunication Company as access network design and otimization senior engineer from June 2006 to August 2015.

\end{IEEEbiography}

\begin{IEEEbiographynophoto}{Mehdi Mirzaeinia}
Biography text here.
\end{IEEEbiographynophoto}


\begin{IEEEbiographynophoto}{Abdelmounaam Rezgui}
Biography text here.
\end{IEEEbiographynophoto}




\end{document}